\documentclass[floatfix,nofootinbib,superscriptaddress,twocolumn,pre,showpacs,showkeys,preprintnumbers]{revtex4-1}
\usepackage{amsmath,amsfonts,amsthm,amssymb}
\usepackage{epsfig,graphics,color,calc,graphicx}
\usepackage{subfigure}
\usepackage{bbm}
\usepackage{mathrsfs}

\newcommand{\bb}[1]{{\mathbb{#1}}}

\newlength{\pecettawidth}
\begin{document}
\title{Stationary uphill currents in locally perturbed Zero Range Processes}

\author{Emilio N.M.\ Cirillo}
\email{emilio.cirillo@uniroma1.it}
\affiliation{Dipartimento di Scienze di Base e Applicate per l'Ingegneria, 
             Sapienza Universit\`a di Roma, 
             via A.\ Scarpa 16, I--00161, Roma, Italy.}

\author{Matteo Colangeli}
\email{matteo.colangeli1@univaq.it}
\affiliation{Dipartimento di Ingegneria e Scienze dell'Informazione e
Matematica, Universit\`a degli Studi dell'Aquila, via Vetoio,
67100 L'Aquila, Italy.}

\begin{abstract}
Uphill currents are observed when mass diffuses in the direction 
of the density gradient. We study 
this phenomenon in stationary conditions in the framework of locally 
perturbed 1D Zero Range Processes (ZRP). 
We show that the onset of currents flowing from the reservoir with smaller density to the one with larger density can be caused by a local asymmetry in the hopping rates on a 
single site at the center of the lattice. 
For fixed injection rates at the boundaries, we prove that  
a suitable tuning of the asymmetry in the bulk may induce uphill diffusion at arbitrarily large, finite volumes. 
We also deduce heuristically 
the hydrodynamic behavior of the model  and connect the local 
asymmetry characterizing the ZRP dynamics to a matching condition relevant for the 
macroscopic problem. 
\end{abstract}




\maketitle

\section{Introduction}
\label{s:introduzione} 
\par\noindent
Fick's law of diffusion stands as one of the basic tenets of the theory of transport phenomena and Irreversible Thermodynamics, and predicts that mass diffuses \textit{against} the density gradient \cite{BSL,Ott}. Nonetheless, there is some increasing experimental and theoretical evidence, in the literature, of diffusive currents flowing from a reservoir with lower density towards one with larger density, that are hence said to go \emph{uphill} \cite{Erleb,Ramirez,Lauer,Sato}.
Such ``anomalous'' currents have been observed and studied in 
different contexts. Consider, for instance, a system made of particles of a certain species $A$, whose diffusive motion obeys the standard Fick's law, namely the current of particles $A$ includes a term proportional to \textit{minus} the density gradient of $A$ itself. Suppose that a second species $B$ is then added, whose interaction with $A$ affects the diffusive motion of the particles of the first species. Thus, a second contribution to the current of particles $A$ arises, related to the density gradient of $B$, that may counterbalance the first contribution. As a result, at variance with the standard Fick's law prescription \cite{KRnjc2016,Kcsr2015}, the species $A$ undergoes a process of uphill diffusion induced by the external potential generated by the species $B$: this is, essentially, the phenomenon highlighted in the seminal paper by Darken \cite{Daime1949}, reporting an experiment of transient diffusion of carbon atoms subjected to a repulsive interaction with silicon particles in a welded specimen, where the silicon content is concentrated on the left of the weld (and negligible on the right).


A second stationary mechanism which is known to produce uphill currents is
related to the presence of a phase transition in non--equilibrium conditions \cite{DPT,CDMPplA2016}.
This phenomenon has been observed in computer simulations in a 
model constituted by a single species undergoing a 
liquid--vapor phase transition. This system, with one boundary 
fixed at the density of the metastable vapor phase and the other 
at the density of the metastable liquid phase, 
exhibits a stationary state in which the current flows from the 
vapor boundary to the liquid one. In particular, in Ref. \cite{CDMPjsp2017} the authors prove the existence of the uphill diffusion phenomenon for a stochastic cellular automaton, in which the particles are subjected to an exclusion rule (preventing the simultaneous presence of two particles with same velocity on a same site) and to a long--range Kac potential \cite{Pres}.
As distinct from the Darken experiment, the mechanism responsible, in this case, for the breaking of the standard diffusive behavior is the creation of a sharp interface located near one of the two boundaries -- called \textit{bump} therein -- separating the vapor and the liquid phases. The density profile results essentially decreasing almost everywhere along the 1D spatial domain, except at the transition region: in fact, the stationary current proceeds downhill in most of the space, but it goes uphill right along the interface.

Noticeably, the occurrence of stationary uphill currents induced by a phase transition was also recently reported in \cite{CGGV}, for a 2D Ising model in contact with two infinite reservoirs fixing the values of the density at the horizontal boundaries.

In this paper we study a different mechanism to 
produce uphill currents, based on a local perturbation of 
a stationary state. The model discussed below allows to recover some of the important features of the physical examples of uphill diffusion mentioned above. In fact, despite being simple enough to permit an analytical solution, it gives rise to a stationary uphill diffusion which is not induced by a phase transition as in \cite{CDMPplA2016,CDMPjsp2017,CGGV}, but is triggered by a local asymmetry in the hopping rates that rule the microscopic dynamics in the bulk. The asymmetry at the center of the lattice stands as a {\em caricature} of the external potential exerted by the silicon particles on the carbon atoms, as described in the Darken experiment \cite{Daime1949}, cf.\ also the set--up discussed 
in Sec.\ III of \cite{CDMP2017}.

The effect of local perturbations of stationary states is 
a fascinating problem which deserved a lot of 
attention in the recent physics and mathematical literature, see e.g. the review \cite{BpA2006}.
A classical question in this field is the so--called \emph{blockage} 
problem, posed in \cite{JL1994} for the totally asymmetric 
simple exclusion process on a ring. The question is whether 
slowing down a single bond on the lattice can ultimately affect the value of the stationary 
current in the infinite volume limit, see 
also \cite{CCMpre2016,SLM15,CMKS2016} for related results for different models.

Differently from the blockage problem, the question addressed in this paper concerns the effect of a local asymmetry in a globally symmetric model. Consider the stationary state of a 1D 
system with 
symmetric dynamics and suppose that a nonvanishing current exists 
due to the coupling of the system with two particle reservoirs at the boundaries. What happens if the dynamics is perturbed and made asymmetric just on 
a single site of the lattice? Is such a local asymmetry 
effective enough to reverse the natural current flowing direction?

More precisely, the model we shall consider is a 1D channel with open boundaries at its extremities (hereafter called ZRP--OB), in contact with two reservoirs. The reservoirs are equipped with assigned particle densities, which also fix the injection rates at the boundaries. 
The dynamics in the channel is symmetric, therefore in the steady state a particle current exists which moves from the reservoir with larger density to the one with smaller density, as prescribed by the Fick's law. Then, on a single site at the center of the lattice, 
the dynamics is modified in such a way that particles locally hop with higher rate
towards the reservoir with larger density.  
More general inhomogeneous random ZRP have been considered in the recent literature \cite{GCS08,GL12}.
We prove that such a {\em bias} may give rise to stationary uphill currents 
in the channel.
In particular, we prove that for any fixed difference 
between the two injection rates it is always possible 
to tune the local asymmetry in order to observe 
an uphill current for arbitrarily large finite volumes. The mechanism is the following: for sufficiently large volumes the density at the boundaries of the channel depends only on the injection rates and not on the local bias; moreover, if the bias is large enough the current changes sign so that the particles move uphill. 
The model we shall use is a 1D Zero Range Process.
More detailed results will be derived by establishing an appropriate form for the intensity 
function, namely the rate at which a site is updated, and eventually this will be chosen proportional 
to the number of particles occupying the site. 
In this case, we shall also develop a heuristic 
argument to derive the hydrodynamics equations. These will be endowed with two matching conditions -- one concerning the 
density function and another its first space derivative --  
at the center of the slab, stemming from the local asymmetry in the hopping rates characterizing the microscopic dynamics.  
We will then solve the problem via a Fourier series expansion and 
we shall finally compare, finding a perfect match, the solution of 
the hydrodynamic problem with the evolution of the original ZRP.  
We also mention that uphill currents are observed in 
queuing network models.

Moreover, we will introduce a periodic version of the inhomogeneous ZRP, in which the channel is coupled at its extremities with two slow sites -- mimicking two finite particle reservoirs --
which can also exchange particle between themselves: the whole 
system thus constitutes a closed circuit (hereafter called ZRP--CC). 
One of the open questions posed in \cite{CDMPjsp2017}, in the context of stochastic particle systems, was the conjectured existence of stationary states with nonvanishing self--sustained currents running in circuits, this phenomenon being also called ``\textit{time crystals}'' in the literature \cite{W,SW}. We shall not tackle rigorously the existence of those fascinating rotating states here; rather, we aim to give theoretical and numerical evidence that the local asymmetry introduced in the ZRP--CC may lead to a stationary state in which the densities of the finite reservoirs are different and the current flows, in the channel, from the reservoir with lower density to the one with larger density (as it was also the case for the ZRP--OB). 
Steady states for ZRP with periodic boundary conditions and spatially varying hopping rates were also discussed in \cite{Ev00,BDL10}.

The paper is organized as follows. In Section~\ref{s:modello} we
introduce the two ZRP models, the ZRP--OB and the ZRP--CC, we define the stationary current and also recall some useful properties.
In Section~\ref{s:uphill} we prove the existence of uphill currents for the ZRP--OB model. 
Section~\ref{s:circuito} is devoted to the study of uphill currents 
for the ZRP--CC. 
In Section~\ref{s:idro} we discuss heuristically the hydrodynamic 
limit of the ZRP--OB and compare the solution of the 
hydrodynamic equation to the profile evolving according to the 
stochastic ZRP dynamics. 
Finaly, Section~\ref{s:conclusioni} is devoted to our 
brief conclusions. 

\begin{figure*}
\centerline{%
{\includegraphics[width=.95\textwidth]{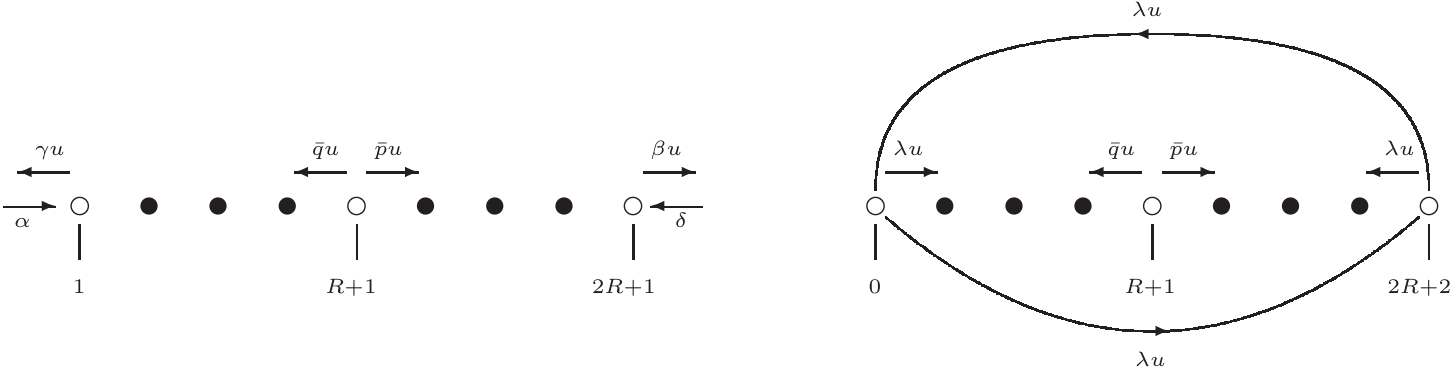}}
}
\caption{Schematic representation of the ZRP--OB model
(left panel) and the ZRP--CC model (right panel). Rates associated with the 
black sites are $pu$ towards the left and 
$qu$ towards the right.}
\label{fig01}
\end{figure*}

\section{The model}
\label{s:modello}
\par\noindent
We define the two ZRP models to be studied in the following sections, see also \cite{S1970,DMP1991,EH2005} for a survey on ZRP models.

\subsection{The ZRP--OB}
\label{s:aperto}
\par\noindent
We consider a positive integer $R$ and define a ZRP on the finite lattice
$\Lambda=\{1,\dots,2R+1\}\subset\mathbb{Z}$.
We consider the finite \emph{state} or \emph{configuration space}
$\Omega_R=\bb{N}^\Lambda$.
Given
$n=(n_1,\dots,n_{2R+1})\in\Omega_R$
the non--negative integer $n_x$ is called \emph{number of particles}
at the site $x\in\Lambda$ in the \emph{state} or \emph{configuration}
$n$.
We let $u:\bb{N}\to\bb{R}_+$, a positive and 
non--decreasing function such that $u(0)=0$, be the \emph{intensity}.
Given $n\in\Omega_R$ such that $n_x>0$ for some $x=1,\dots,2R+1$, 
we let $n^{x,x\pm1}$ be
the configuration obtained by moving a particle from the site $x$ 
to the site $x\pm1$; in particular, we understand 
$n^{1,0}$ and  
$n^{2R+1,2R+2}$ to be the configurations obtained by removing a particle from the site, respectively, $1$, 
 and $2R+1$.
Similarly, 
we denote by $n^{0,1}$ and $n^{2R+2,2R+1}$ the configurations 
obtained by adding a particle to the site $1$ and $2R+1$, 
respectively. 

Given $p,q,\bar{p},\bar{q},\alpha,\beta,\gamma,\delta>0$ we
set 
$q_1=\gamma$,
$q_x=q$ for $x=2,\dots,R$ and $x=R+2,\dots,2R+1$, $q_{R+1}=\bar{q}$,
$p_x=p$ for $x=1,\dots,R$ and $x=R+2,\dots,2R$, $p_{R+1}=\bar{p}$, 
and $p_{2R+1}=\beta$. 

We then consider the ZRP--OB model, defined as 
the continuous time Markov jump process $n(t)\in\Omega_R$, $t\ge0$,
with rates
\begin{equation}
\label{ape010}
r(n,n^{0,1})=\alpha
\;\textrm{ and }\;
r(n,n^{2R+2,2R+1})=\delta
\end{equation}
for particles injection at the boundaries,
and with rates
\begin{equation}
\label{ape025}
r(n,n^{x,x-1})=q_xu(n_x)
\textrm{ for } x=1,\dots,2R+1
\end{equation}
for bulk leftwards displacements, and
\begin{equation}
\label{ape028}
r(n,n^{x,x+1})=p_xu(n_x)
\textrm{ for } x=1,\dots,2R+1
\end{equation}
for bulk rightwards displacements
(see Figure~\ref{fig01}).
Note that equations \eqref{ape025} and \eqref{ape028} 
for $x=1$ and $x=2R+1$, respectively, account 
for the particles removal at the boundaries.
The generator of the dynamics can be written as
\begin{equation}
\label{ape100}
\begin{array}{rcl}
(L_Rf)(n)
&\!\!=&\!\!
 \alpha(f(n^{0,1})-f(n))
\\
&&\!\!
{\displaystyle
 +
 \!\!\!\sum_{x=1}^{2R+1}
 [q_xu(n_x)(f(n^{x,x-1})-f(n))
}
\\
&&\!\!
{\displaystyle
\phantom{
 +
 \!\!\!\sum_{x=1}^{2R+1}
 [
        }
 +p_xu(n_x)(f(n^{x,x+1})-f(n))]
}
\\
&&\!\!
 +
 \delta(f(n^{2R+2,2R+1})-f(n))
\end{array}
\end{equation}
for any real function $f$ on $\Omega_R$.

This means that particles hop almost everywhere on the lattice to the neighboring sites with rates 
$qu(n_x)$ and $pu(n_x)$. 
At the center of the lattice, instead, different rates are assumed, namely
$\bar{q}u(n_x)$ and $\bar{p}u(n_x)$.
The system is ``open'' in the sense that a particle hopping from the sites 
$1$ or $2R+1$ can leave the channel via, respectively, a left 
or a right move, with rates $\gamma u(n_1)$ and $\beta u(n_{2R+1})$.
Finally, particles are injected in the channel at the left and right 
boundaries with rates, respectively, $\alpha$ and $\delta$.  

No further characterization of the (infinite) reservoirs is required for the ZRP--OB, as the action of the reservoirs is suitably described in terms of the injection rates $\alpha$ and $\delta$. Nevertheless, it may be useful to think of each injection rate as being proportional to the (fixed, for the ZRP--OB model) particle density of the corresponding reservoir, as proposed in \cite{CGGV} for a continuous--time dynamics, see also \cite{CDMPplA2016,CDMPjsp2017} in the case of a cellular automaton. Hence, a larger injection rate corresponds to a larger density of the reservoir.

\subsection{The ZRP--CC}
\label{s:periodico}
\par\noindent
The definition of the model is similar to the ZRP--OB.
We consider the positive integers $R,N$ and define a ZRP on the finite 
torus $\Lambda=\{0,1,\dots,2R+2\}\subset\mathbb{Z}$.
We consider the finite \emph{configuration space}
$\Omega_{R,N}=
\{
n\in\{0,\dots,N\}^\Lambda,\,
\sum_{x\in\Lambda}n_x=N
\}
$.
Given
$n=(n_0,\dots,n_{2R+2})\in\Omega_{R,N}$
the non--negative integer $n_x$ is called \emph{number of particles}
at the site $x\in\Lambda$ in the \emph{configuration}
$n$.
We let $u:\bb{N}\to\bb{R}_+$, a positive and 
non--decreasing function such that $u(0)=0$, be the \emph{intensity}.
Given $n\in\Omega_R$ such that $n_x>0$ for some $x=0,\dots,2R+2$, 
we let $n^{x,x\pm1}$ be
the configuration obtained by moving a particle from the site $x$ 
to the site $x\pm1$, where we denote by 
$n^{0,-1}$ the configuration obtained by moving a particle from the site $0$
to the site $2R+2$, 
and  
by $n^{2R+2,2R+3}$ the configuration obtained by moving a particle from 
the site $2R+2$ to the site $0$.

Given $p,q,\bar{p},\bar{q},\lambda>0$ we
set 
$q_x=q$ for $x=1,\dots,R$ and $x=R+2,\dots,2R+1$, $q_{R+1}=\bar{q}$,
$p_x=p$ for $x=1,\dots,R$ and $x=R+2,\dots,2R+1$, and $p_{R+1}=\bar{p}$.
We consider the \emph{periodic} ZRP defined as 
the continuous time Markov jump process $n(t)\in\Omega_R$, $t\ge0$,
with rates
\begin{equation}
\label{per010}
r(n,n^{0,\pm1})=\lambda u(n_0)
\end{equation}
and
\begin{equation}
\label{per015}
r(n,n^{2R+2,2R+2\pm1})=\lambda u(n_{2R+2})
\end{equation}
for the boundary conditions, and with rates
\begin{equation}
\label{per025}
r(n,n^{x,x-1})=q_xu(n_x)
\textrm{ for }
x=1,\dots,2R+1 
\end{equation}
for bulk leftwards displacements, and
\begin{equation}
\label{per028}
r(n,n^{x,x+1})=p_xu(n_x)
\textrm{ for }
x=1,\dots,2R+1 
\end{equation}
for bulk rightwards displacements
(see Figure~\ref{fig01}).
The generator of the dynamics can be written as 
\begin{equation}
\label{per100}
\begin{array}{rcl}
(L_{R,N}f)(n)
&\!\!=&\!\!
 \lambda u(n_0)(f(n^{0,-1})-f(n))
\\
&&\!\!
 +
 \lambda u(n_0)(f(n^{0,1})-f(n))
\\
&&\!\!
{\displaystyle
 +
 \!\!\!
 \sum_{x=1}^{2R+1}
 [q_xu(n_x)(f(n^{x,x-1})-f(n))
}
\\
&&\!\!
{\displaystyle
 \phantom{
 +
 \!\!\!
 \sum_{x=1}^{2R+1}
 [
 }
 +p_xu(n_x)(f(n^{x,x+1})-f(n))]
}
\\
&&\!\!
 +
 \lambda u(n_{2R+2})(f(n^{2R+2,2R+1})-f(n))
\\
&&\!\!
 +
 \lambda u(n_{2R+2})(f(n^{2R+2,2R+3})-f(n))
\end{array}
\end{equation}
for any real function $f$ on $\Omega_R$.

The ZRP--CC model differs from the ZRP--OB for the 
boundary conditions: particles can neither exit nor enter the system. 
Furthermore, the sites $0$ and $2R+2$ are updated with rates proportional to 
$\lambda$.
The interesting case, from the modelling perspective, is that in which $\lambda$ 
is much smaller than one: namely, the boundary sites are slowed down
and mimic the action of large particle reservoirs.

\begin{figure*}
\centerline{%
{\includegraphics[width=.8\textwidth]{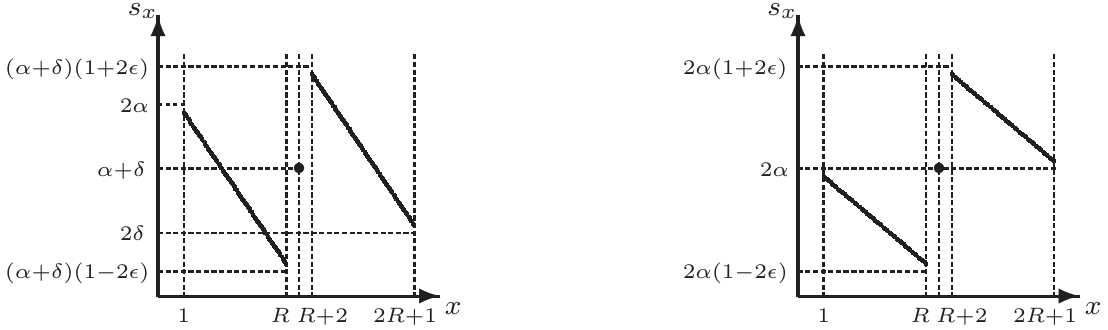}}
}
\caption{Fugacity profile, for $R$ large, in the case $\alpha>\delta$ (left panel) 
and $\alpha=\delta$ (right panel).}
\label{fig02}
\end{figure*}

\begin{figure*}
\centerline{%
{\includegraphics[width=.95\textwidth]{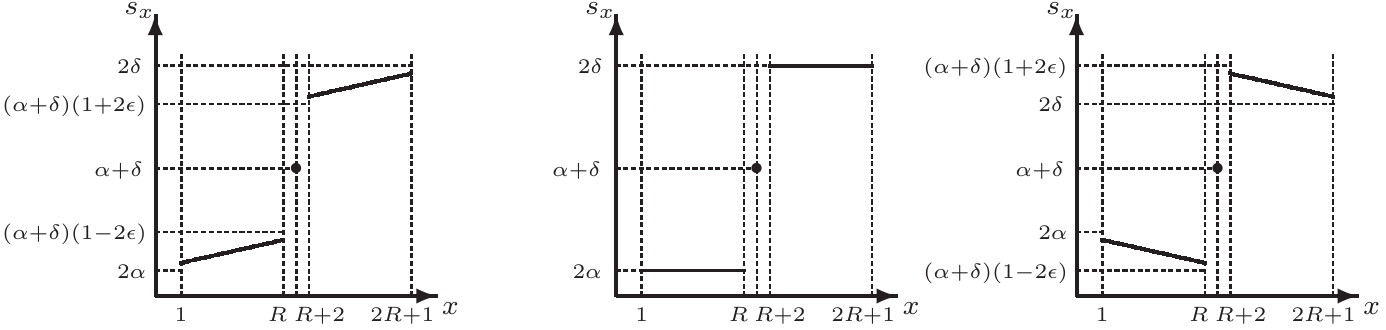}}
}
\caption{Fugacity profile, for $R$ large, in the case $\alpha<\delta$ 
for 
$\epsilon<\epsilon_\textrm{c}$ (left panel), 
$\epsilon=\epsilon_\textrm{c}$ (central panel), and 
$\epsilon_\textrm{c}<\epsilon$ (right panel).} 
\label{fig03}
\end{figure*}

\section{Uphill currents in the ZRP--OB}
\label{s:uphill} 
\par\noindent
In this section we shall prove that the ZRP--OB can 
exhibit stationary uphill currents. More precisely, we shall consider the 
process described by the generator given in \eqref{ape100}, and show that, for a particular 
choice of the parameters, the steady state is characterized by a current flowing
from the reservoir with smaller density to the one with larger density.  

\subsection{Stationary measure for the ZRP--OB}
\label{s:ap-sta} 
\par\noindent
Consider the ZRP--OB defined in Section~\ref{s:aperto}.
A probability measure $\mu_R$ on $\Omega_R$ is stationary for the ZRP 
if and only if 
\begin{equation}
\label{uph000}
\sum_{n\in\Omega_R}\mu_R(n)(L_Rf)(n)=0
\end{equation}
for any function $f$.  
A sufficient condition is provided by the \emph{balance equation}
\begin{equation}
\label{uph010}
\sum_{m\neq n}\mu_R(m)r(m,n)
=
\mu_R(n)\sum_{m\neq n}r(n,m)
\end{equation}
for any $n\in\Omega_R$.

Consider the positive reals $s_1,\dots,s_{2R+1}$,
called \emph{fugacities},
and the 
product measure on the space $\Omega_R$
defined as 
\begin{equation}
\label{uph020}
\nu_R(n)=\prod_{x=1}^{2R+1}\nu_x(n_x)
\;\;\textrm{ with }\;\;
\nu_x(n_x)=\frac{1}{Z_x}\frac{s_x^{n_x}}{u_x(n_x)!}
\end{equation}
where
$u_x(k)!=1$ if $k=0$
and 
$u_x(k)!=u_x(1)\cdots u_x(k)$ if $k\ge1$
and 
\begin{equation}
\label{uph040}
Z_x=\sum_{k=0}^\infty \frac{s_x^{k}}{u_x(k)!}
\;.
\end{equation}

By exploiting equation \eqref{uph010},
or by applying \eqref{uph000} to the functions $f(n)=n_x$ for any $x$, 
it can be proven that 
$\nu$ is stationary for the ZRP--OB provided the reals 
$s_x$ satisfy the following equations:
\begin{equation}
\label{uph050}
\begin{array}{rcl}
(\gamma+p)s_1
&\!\!=&\!\!
\alpha+qs_2
\\
(q+p)s_x
&\!\!=&\!\!
ps_{x-1}+qs_{x+1}
\;\;\textrm{ for }
x=2,\dots,R-1
\\
(q+p)s_R
&\!\!=&\!\!
ps_{R-1}+\bar{q}s_{R+1}
\\
(\bar{q}+\bar{p})s_{R+1}
&\!\!=&\!\!
ps_{R}+qs_{R+2}
\\
(q+p)s_{R+2}
&\!\!=&\!\!
\bar{p}s_{R+1}+qs_{R+3}
\\
(q+p)s_x
&\!\!=&\!\!
ps_{x-1}+qs_{x+1}
\;\;\textrm{ for }
x=R+3,\dots,2R
\\
(q+\beta)s_{2R+1}
&\!\!=&\!\!
ps_{2R}+\delta
\\
\end{array}
\end{equation}
After some simple algebra we get the equations
\begin{equation}
\label{uph060}
\begin{array}{rcl}
ps_{R}-\bar{q}s_{R+1}
&\!\!=&\!\!
ps_{R-1}-qs_{R}
\\
\bar{p}s_{R+1}-qs_{R+2}
&\!\!=&\!\!
ps_{R}-\bar{q}s_{R+1}
\\
ps_{2R}-qs_{2R+1}
&\!\!=&\!\!
\beta s_{2R+1}-\delta
\\
ps_x-qs_{x+1}
&\!\!=&\!\!
\alpha-\gamma s_1
\\
\end{array}
\end{equation}
for
$x=1,\dots,R-1$
and 
$x=R+2,\dots,2R$,
which reduce to 
\cite[equation~(13)]{LMS2005}
in the case $(\bar{q},\bar{p})=(q,p)$.
These equations admits a unique solution to be discussed 
in detail in the sequel for a particular choice of the parameters 
$p,q,\bar{p},\bar{q},\alpha,\beta,\gamma$, and $\delta$.

\subsection{Stationary current and density profile for the ZRP--OB}
\label{s:ap-cur} 
\par\noindent
The main quantities of interest, in our study, are the stationary 
\emph{density} or \emph{occupation number} profiles
\begin{equation}
\label{uph070}
\rho_x
=
\nu_x[n_x]
=
\frac{1}{Z_x}\sum_{k=1}^\infty k\,\frac{s_x^k}{u(k)!}
=
s_x\frac{\partial}{\partial s_x}\log Z_x
\end{equation}
see \cite{CCM2016MMS} for the details, and 
the stationary \emph{current}
\begin{equation}
\label{uph080}
\begin{array}{rcl}
J_{R,x}
&\!=&\!
\nu_R[u(n_x)p_x-u(n_{x+1})q_{x+1}]
\\
&\!=&\!
p_x\nu_x[u(n_x)]-q_{x+1}\nu_{x+1}[u(n_{x+1})]
\\
&\!=&\!
p_xs_x-q_{x+1}s_{x+1}
\end{array}
\end{equation}
for $x=1,\dots,2R$,
where we omitted the last straightforward computation.
The stationary current 
represents the difference between the
average number of particles crossing a bond between two adjacent
sites on the lattice from the left to the right and the corresponding number in the opposite direction.
Equations \eqref{uph060} shows that the stationary current does not depend 
on the site $x$, therefore we shall simply write $J_R\equiv J_{R,x}$.

Note that it was possible to express the current in terms 
of the fugacities without relying on any specific choice for the 
intensity function $u$. Yet, an explicit form for $u$ is needed in the computation of the density profile. In general 
it can be proven, see \cite{CCM2016MMS}, that
\begin{equation}
\label{uph090}
\frac{\partial\rho_x}{\partial s_x}
=
\frac{1}{s_x}(\nu_x[n_x^2]-(\nu_x[n_x])^2)>0
\;\;,
\end{equation}
hence, at each site, the stationary mean occupation number is an increasing 
function of the local fugacity.

Particularly relevant cases are the so--called \textit{independent particle} 
and the \textit{simple exclusion--like} ZRP models, in which 
the intensity function is respectively given by $u(k)=k$ and 
$u(k)=1$ for $k\ge1$ (recall that $u(0)=0$). In these two cases it is 
easy to prove that 
$Z_x^\textrm{ip}=\exp\{s_x\}$ and 
$Z_x^\textrm{se}=1/(1-s_x)$ for $s_x<1$, respectively. 
Hence, by \eqref{uph070}, one has 
\begin{equation}
\label{uph095}
\rho_x^\textrm{ip}=s_x
\;\;\textrm{ and }\;\;
\rho_x^\textrm{se}=\frac{s_x}{1-s_x} \textrm{ for } s_x<1
\end{equation} 
for the independent particle and the simple exclusion--like models, 
respectively.

\begin{figure*}
\centerline{%
\hspace{.2 cm}
{\includegraphics[width=0.55\textwidth]{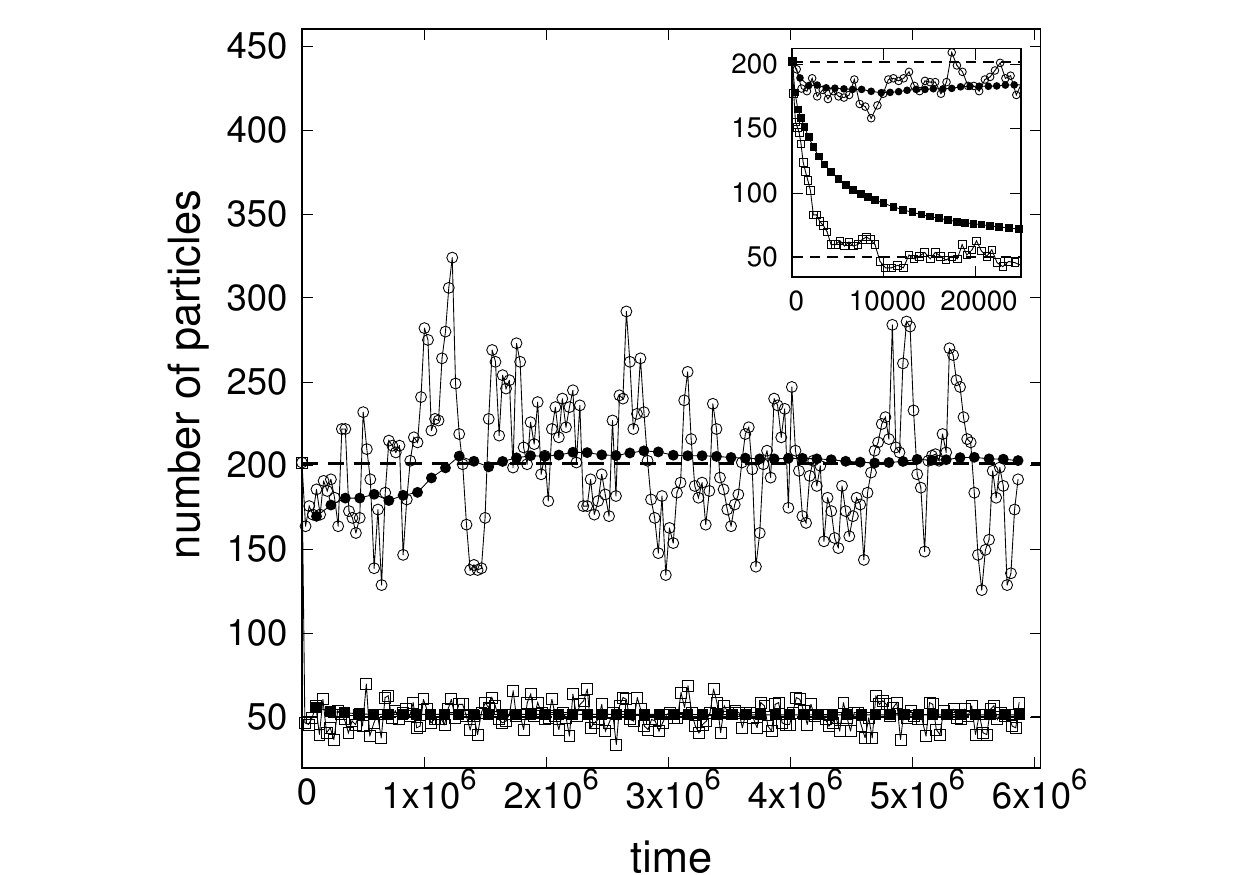}}
\hspace{-1.1 cm}
{\includegraphics[width=0.55\textwidth]{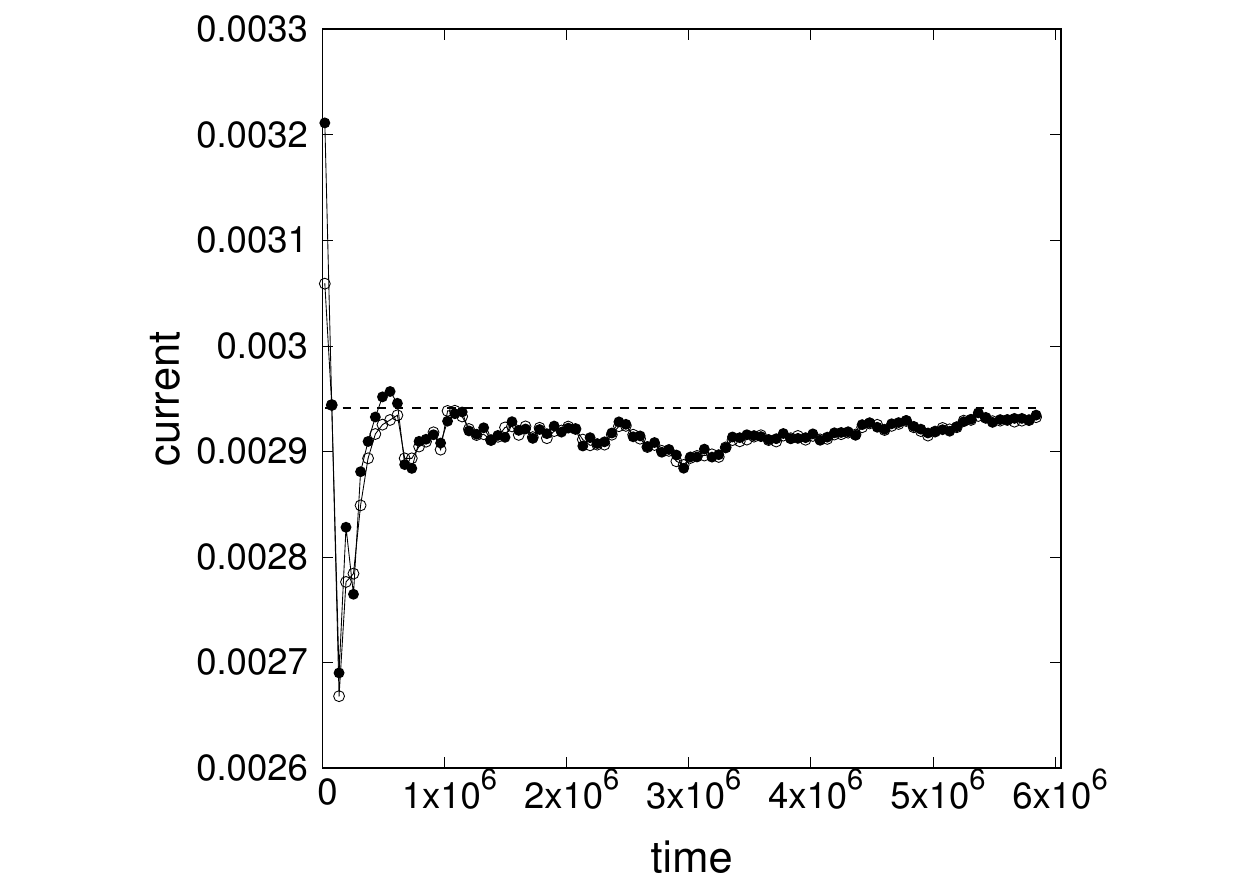}}
}
\caption{Monte Carlo results for the almost everywhere symmetric ZRP--OB with $R=50$, $\epsilon=0.4$, $\alpha=0.2$, $\delta=0.3$. The initial datum used in the simulations is a uniform configuration with $2$ particles per site.
Squares and circles refer, respectively, to the independent particle and
the simple exclusion--like models. 
The dashed lines represent the exact solution. 
Left panel: the total number of particles in the system is measured 
vs.\ time. The open symbols are the
instantaneous values and the solid symbols are the time--averaged values. Time is measured from the beginning of the dynamics. The inset in the upper right corner is a magnification of the larger figure at short times, and shows the rapid relaxation of the time--averaged total number of particles in the independent particle case to the theoretical value given in \eqref{uph135}.
Right panel: particle currents measured at the boundaries vs.\ time.
Open and solid symbols refer, respectively, to the left and the right boundary. Time is counted starting from the ``thermalization'' time $2\times10^6$, taken as the origin of the horizontal axis.
}
\label{fig04}%
\end{figure*}

\subsection{The almost everywhere symmetric ZRP--OB}
\label{s:ap-asy} 
\par\noindent
We shall fix 
$q=p=\gamma=\beta=1/2$,
$\bar{q}=1/2-\epsilon$, and 
$\bar{p}=1/2+\epsilon$ for some $\epsilon\in[0,1/2)$.
In this case, the solution of the equations \eqref{uph060} is linear 
in $x$ and can be written as 
\begin{equation}
\label{uph100}
s_x=
\left\{
\begin{array}{ll}
\!\!
\sigma x+2\alpha 
& \textrm{ for }x=1,\dots,R
\\
\!\!
\alpha+\delta 
& \textrm{ for }x=R+1
\\
\!\!
\sigma x+2\alpha +4\epsilon(\alpha+\delta) 
& \textrm{ for }x=R+2,\dots,2R+1
\\
\end{array}
\right.
\end{equation}
with slope 
\begin{equation}
\label{uph110}
\sigma=\frac{1}{R+1}[\delta-\alpha-2\epsilon(\alpha+\delta)]
\;\;.
\end{equation}
To draw the fugacity profiles $s_x$ for $x=1,\dots,2R+1$, 
shown in Figures~\ref{fig02} and \ref{fig03}, 
it is useful to compute 
\begin{equation}
\label{uph115}
\begin{array}{l}
s_1=\sigma+2\alpha
\\
s_R=-\sigma+(\alpha+\delta)(1-2\epsilon)
\\
s_{R+2}=\sigma+(\alpha+\delta)(1+2\epsilon)
\\
s_{2R+1}=-\sigma+2\delta
\,.
\end{array}
\end{equation}

In the case $\alpha>\delta$, 
recalling that $\epsilon\in[0,1/2)$, 
we have that $\sigma<0$,
$2\alpha>\alpha+\delta>2\delta$,
$0<(\alpha+\delta)(1-2\epsilon)\le\alpha+\delta$,
and
$\alpha+\delta\le(\alpha+\delta)(1+2\epsilon)<2(\alpha+\delta)$.
This explains the graph shown in the left panel of Figure~\ref{fig02},
portraying the fugacity profiles 
for $R$ large, in which case the terms $\pm\sigma$ in \eqref{uph115} 
are small.

In the case $\alpha=\delta$, 
recalling that $\epsilon\in[0,1/2)$, 
we have that $\sigma=-4\epsilon\alpha/(R+1)<0$
and 
$0<2\alpha(1-2\epsilon)\le2\alpha\le2\alpha(1+2\epsilon)$.
This case is shown in the right panel of Figure~\ref{fig02}, where the terms $\pm\sigma$ are still assumed to be small.

The discussion of the the case $\alpha<\delta$ is more delicate, since
the sign of $\sigma$ depends on the value of the difference $\delta-\alpha$. 
More precisely, it holds
\begin{equation}
\label{uph120}
\sigma\ge0
\;\;\textrm{ if and only if }\;\;
\epsilon\le
\frac{1}{2}\frac{\delta-\alpha}{\alpha+\delta}
\equiv
\epsilon_\textrm{c}\;,
\end{equation}
where we have introduced the \emph{critical bias} 
$\epsilon_\textrm{c}$. Thus, we have to distinguish three cases. 

For $0\le\epsilon<\epsilon_\textrm{c}$ we have that 
$\sigma>0$,
$2\alpha<(\alpha+\delta)(1-2\epsilon)\le\alpha+\delta$,
and
$\alpha+\delta\le(\alpha+\delta)(1+2\epsilon)<2\delta$.
For $\epsilon=\epsilon_\textrm{c}$ we have that 
$\sigma=0$, 
$(\alpha+\delta)(1-2\epsilon_\textrm{c})=2\alpha$, 
and 
$(\alpha+\delta)(1+2\epsilon_\textrm{c})=2\delta$.
For $\epsilon_\textrm{c}<\epsilon<1/2$ we have that 
$\sigma<0$,
$2\alpha>(\alpha+\delta)(1-2\epsilon)>0$,
and
$2(\alpha+\delta)>(\alpha+\delta)(1+2\epsilon)>2\delta$.
The graphs in Figure~\ref{fig03} represent the fugacity profiles 
for $R$ large, in the three cases.\\
Remarkably, the presence of a critical value for the {\em bias}, marking the transition from a regime of standard (downhill) diffusion to another regimes of uphill diffusion, was also reported in \cite{CDMP2017}, cf. Figure 4 therein. In that work, much in the same spirit of the ZRP--OB model, the diffusion of particles in the channel results from the balance between the standard diffusive behavior induced by the reservoirs and the uphill motion triggered by the Kac potential in the bulk (whose effect is only visible in a neighborhood around the central site of the lattice).

As already mentioned in Section~\ref{s:ap-cur}, the stationary current
can be computed from the knowledge of the fugacity profile without 
specifying the intensity function. Applying \eqref{uph080} and 
\eqref{uph100} we find 
\begin{equation}
\label{uph130}
J_R
=-\frac{1}{2}\sigma
=-\frac{1}{2(R+1)}[\delta-\alpha-2\epsilon(\alpha+\delta)]
\;\;.
\end{equation} 
On the other hand, 
to compute the density profile
it is necessary to consider a particular form for the intensity function, 
see \eqref{uph095} for the independent particle and the simple exclusion--like 
cases. 

For the independent particle model, the fugacity profiles 
shown in Figures~\ref{fig02} and \ref{fig03} 
correspond to the density profiles. 
In particular, by summing up the density profile $\rho_x$ for 
$x=1,\dots,2R+1$, we find the average total number of particles 
in the channel in the steady state:
\begin{equation}
\label{uph135}
N^\textrm{ip}=(\alpha+\delta)(2R+1)
\;\;.
\end{equation}
Moreover, 
in the case $\alpha>\delta$, $\sigma<0$ implies that $J_R>0$. The 
current goes downhill, i.e. it flows from the reservoir with larger 
density (characterized by the injection rate $\alpha$) towards the reservoir with smaller density (with injection rate $\delta$). \\
When $\alpha=\delta$, $\sigma<0$ implies that $J_R>0$. The 
diffusion is now uphill: indeed, in spite of the equality of the injection rates, the current goes from the boundary site $1$,
with lower density, to the site $2R+1$, with higher density. The effect, though, is barely visible because 
$s_{2R+1}-s_1=8\epsilon\alpha/(R+1)$ vanishes for $R$ large. 
It is also interesting to note that, for $R$ sufficiently large, the density profile corresponding to the case $\alpha=\delta$ recovers, qualitatively, the plot portrayed in Figure 2 of \cite{CDMP2017}, referring to a scenario similar to the one considered here for the ZRP--OB model, where the injection rates at the boundaries coincide.\\
In the case $\alpha<\delta$, finally, 
for $\epsilon<\epsilon_\textrm{c}$ the diffusion is downhill, 
for $\epsilon=\epsilon_\textrm{c}$ the current vanishes, 
and 
for $\epsilon_\textrm{c}<\epsilon$ the diffusion is uphill.

We cannot write a general formula for the density profile for any 
choice of the intensity function $u$. But, using  
\eqref{uph070}, we have that $\rho_{x+1}>\rho_x$ if and only 
if $s_{x+1}>s_x$. This implies that the results we deduced 
for the independent particle model are, indeed, completely general. 

Let us now compare our exact results with the Monte Carlo simulations. 
The model has been 
simulated as follows: call $n$ the
configuration at time $t$, then
(i) a number $\tau$ is picked up at random with
exponential distribution of
parameter $U=\alpha+\delta+\sum_{x=1}^{2R+1}u_x(n_x)$ and time is update to
$t+\tau$; 
(ii)
an integer $y$ in $0,1,\dots,2R+2$ is chosen at random on the lattice 
with probability
$\pi_y=u_y(n_y)/U$
for $y=1,\dots,2R+1$, 
$\pi_0=\alpha/U$, 
and 
$\pi_{2R+2}=\delta/U$ (note that, for simplicity, we skipped 
the time dependence in the notation);
(iii) 
if $y\neq0,2R+2$
a particle is moved from the site $y$ to the site $y+1$
with probability $p_y/(q_y+p_y)$ (in the case $y=2R+1$ the particle is 
removed) or to the site $y-1$
with probability $q_y/(q_y+p_y)$ (in the case $y=1$ the particle is removed), 
if $y=0$ a particle is added to the site $1$, 
if $y=2R+2$ a particle is added to the site $2R+1$.

\begin{figure*}[t]
\centerline{%
\hspace{.2 cm}
{\includegraphics[width=0.55\textwidth]{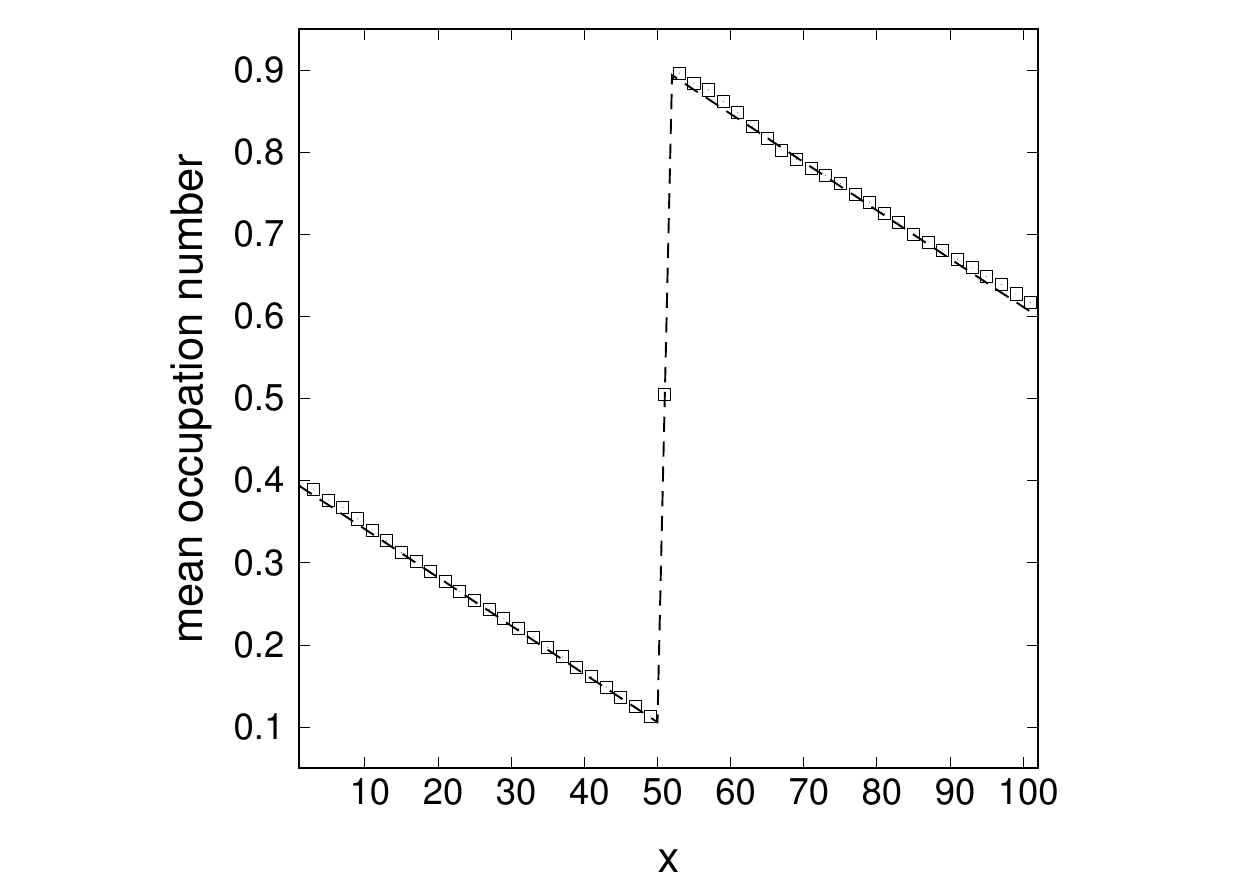}}%
\hspace{-1.1 cm}
{\includegraphics[width=0.55\textwidth]{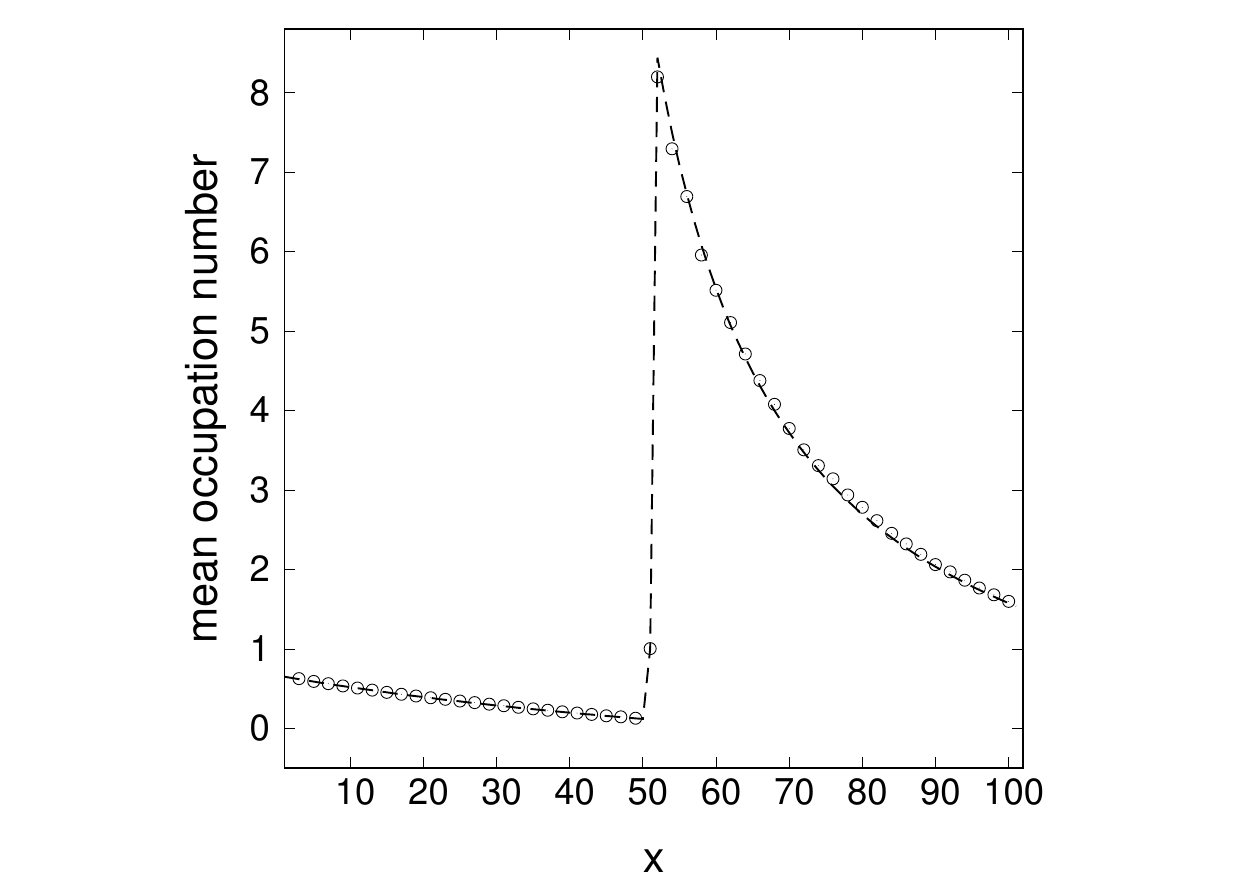}}%
}
\caption{Monte Carlo results for the density profile of 
the almost everywhere symmetric ZRP--OB with $R=50$, $\epsilon=0.4$, $\alpha=0.2$ and $\delta=0.3$.
Squares and circles refer, respectively, to the independent particle 
(left panel) and
the simple exclusion--like (right panel) models. 
The dashed lines represent the exact solution given in \eqref{uph095}. 
The stationary density profiles have been computed by averaging, after the thermalization time is reached, over a set of $10^6$ instantaneous particle configurations.}
\label{fig05}%
\end{figure*}

In Figures~\ref{fig04} we report the Monte Carlo 
measure of the total number of particles in the channel and of the 
boundary currents as functions of time, for the independent particle and the simple exclusion--like models. The values of the 
parameters used in the simulations are indicated in the caption. 
Both panels of Figure ~\ref{fig04} show that when time is large enough the time--averaged values of the total number of particles and of the currents tend to the analytical results \eqref{uph130} and \eqref{uph135} (dashed lines). 
Concerning the theoretical value of the total number of particles, note that the analytic 
expression \eqref{uph135} only applies to the independent particle 
case; in the simple exclusion--like model we summed up
numerically,
for $x=1,\dots,2R+1$, 
the values of $\rho_x$ given by \eqref{uph095}, with 
$s_x$ in \eqref{uph100}.

The data in Figure~\ref{fig04} (left panel) give also an insight into the magnitude of the ``thermalization'' time, namely the time interval in which the time--averaged total number of particles converges to the corresponding theoretical stationary value. As visible in the inset of Figure~\ref{fig04} (left panel), the thermalization time in the independent particle case is considerably smaller than that observed in the simple exclusion--like model, which is of the order of $2\times10^6$ (for the given initial datum used in the simulations). It should also be noted that the steady state fluctuations of the instantaneous total number of particles around the time--averaged value are larger in the simple exclusion--like model. To numerically check the convergence of the current to its theoretical value, see Figure~\ref{fig04} (right panel), we thus skipped the initial transient dynamics and measured the current starting from the time $2\times10^6$.

The same procedure was also adopted for the measure of the stationary density 
profiles, reported in Figure~\ref{fig05}. The match between the 
Monte Carlo numerical measure and the exact results is striking. 
As expected (see \eqref{uph095}), the density 
profile in the simple exclusion--like model is not linear. 
Note, also, that in the simple exclusion--like model 
no symmetry between the left and the right halves of the lattice 
exists. Moreover, though the values of the boundary rates $\alpha$ and $\delta$ 
used in the simulations are the same in the two considered models, completely 
different values of the density at the boundary sites are obtained. This 
suggests, hence, that the dynamics in the bulk significantly affects the value 
of the density at the boundaries.

\section{Uphill currents in the ZRP--CC}
\label{s:circuito} 
\par\noindent
In this section we shall prove that also the ZRP--CC can
exhibit anomalous uphill currents. More precisely, we shall consider the 
process described by the generator given in \eqref{per100} and discuss the effect produced, in the steady state, by the local asymmetry in the bulk and by the two slow boundary sites.

\subsection{Stationary measure for the ZRP--CC}
\label{s:per-sta} 
\par\noindent
For the periodic ZRP introduced in Section~\ref{s:periodico}, the 
invariant measure $\mu_{R,N}$ satisfies
\begin{equation}
\label{cir000}
\sum_{n\in\Omega_{R,N}}\mu_{R,N}(n)(L_{R,N}f)(n)=0\;.
\end{equation}
With arguments similar to those developed in Section~\ref{s:uphill}, 
see also \cite[equation~(15)]{EH2005},
it can be proven 
that the \emph{invariant} or \emph{stationary measure}
of the ZRP--CC process attains the form
\begin{equation}
\label{cir010}
\nu_{R,N}(n)
=
\frac{1}{Z_{R,N}}
\prod_{x=0}^{2R+2}
\frac{s_x^{n_x}}{u(n_x)!}
\end{equation}
for any $n\in\Omega_{R,N}$, where the
\emph{partition function}
$Z_{R,N}$ is the normalization constant 
\begin{equation}
\label{cir015}
Z_{R,N}
=
\sum_{n\in\Omega_{R,N}}
\prod_{x=0}^{2R+2}
\frac{s_x^{n_x}}{u(n_x)!}
\end{equation}
and $s_0,\dots,s_{2R+2}$ are 
not negative real numbers satisfying the following equations
\begin{equation}
\label{cir020}
\begin{array}{rcl}
2\lambda s_0
&\!\!=&\!\!
qs_1+\lambda s_{2R+2}
\\
(q+p)s_1
&\!\!=&\!\!
\lambda s_0+q s_2
\\
(q+p)s_x
&\!\!=&\!\!
ps_{x-1}+qs_{x+1}
\;\;\textrm{ for }
x=2,\dots,R-1
\\
(q+p)s_R
&\!\!=&\!\!
ps_{R-1}+\bar{q}s_{R+1}
\\
(\bar{q}+\bar{p})s_{R+1}
&\!\!=&\!\!
ps_{R}+qs_{R+2}
\\
(q+p)s_{R+2}
&\!\!=&\!\!
\bar{p}s_{R+1}+qs_{R+3}
\\
(q+p)s_x
&\!\!=&\!\!
ps_{x-1}+qs_{x+1}
\;\;\textrm{ for }
x=R+3,\dots,2R
\\
(q+p)s_{2R+1}
&\!\!=&\!\!
ps_{2R}+\lambda s_{2R+2}
\\
\end{array}
\end{equation}
With simple algebra we get the equations
\begin{equation}
\label{cir030}
\begin{array}{rcl}
2\lambda s_0
&\!\!=&\!\!
qs_1+\lambda s_{2R+2}
\\
ps_{R}-\bar{q}s_{R+1}
&\!\!=&\!\!
ps_{R-1}-qs_{R}
\\
\bar{p}s_{R+1}-qs_{R+2}
&\!\!=&\!\!
ps_{R}-\bar{q}s_{R+1}
\\
ps_{2R+1}-\lambda s_{2R+2}
&\!\!=&\!\!
p s_{2R}-q s_{2R+1}
\\
ps_x-qs_{x+1}
&\!\!=&\!\!
\lambda s_0-q s_1
\\
\end{array}
\end{equation}
for 
$x=1,\dots,R-1$
and 
$x=R+2,\dots,2R$.
These equations admits a class of $\infty^1$ solutions 
that 
will be discussed 
in detail in the sequel for a particular choice of the parameters 
$p,q,\bar{p},\bar{q}$, and $\lambda$
defining the rates.

\subsection{Stationary current and density profile for the ZRP--CC}
\label{s:per-cur} 
\par\noindent
We shall focus, again, on the stationary 
\emph{density} profile
\begin{equation}
\label{cir040}
\rho_{R,N,x}
=
\nu_{R,N}[n_x]
=
\frac{1}{Z_{R,N}}\sum_{n\in\Omega_{R,N}}
n_x \prod_{y=0}^{2R+2}\frac{s_x^{n_y}}{u(n_y)!}
\end{equation}
and the stationary \emph{current}
\begin{equation}
\label{cir050}
J_{R,N,x}=
\nu_{R,N}[u(n_x)p_x-u(n_{x+1})q_{x+1}]
\end{equation}
for $x=1,\dots,2R$.
With the same arguments used to prove 
\cite[equation~(11)]{EH2005} we get
\begin{equation}
\label{cir060}
\nu_{R,N}[u(n_x)]
=
\frac{Z_{R,N-1}}{Z_{R,N}}s_x\;,
\end{equation}
hence
\begin{equation}
\label{cir070}
J_{R,N,x}=
\frac{Z_{R,N-1}}{Z_{R,N}}
(p_xs_x-q_{x+1}s_{x+1})
\end{equation}
for $x=1,\dots,2R$,
where the last equalities follows from 
\cite[equation~(11)]{EH2005}. 
Equations \eqref{cir030} proves that the current does not depend 
on the site $x$, hence we shall simply write $J_{R,N}\equiv J_{R,N,x}$.

\begin{figure*}
\centerline{%
\hspace{-1.2 cm}
{\includegraphics[width=0.5\textwidth]{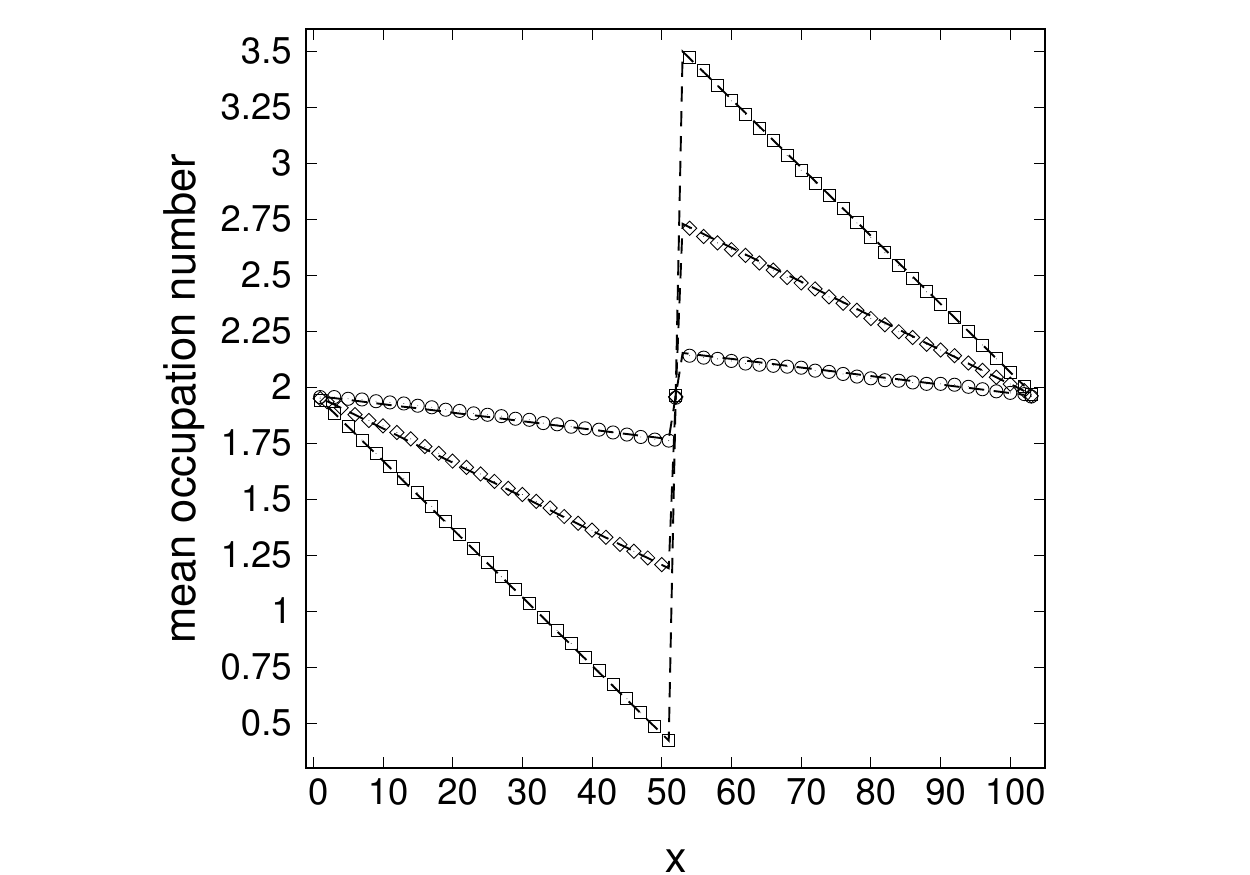}}%
\hspace{0. cm}
{\includegraphics[width=0.5\textwidth]{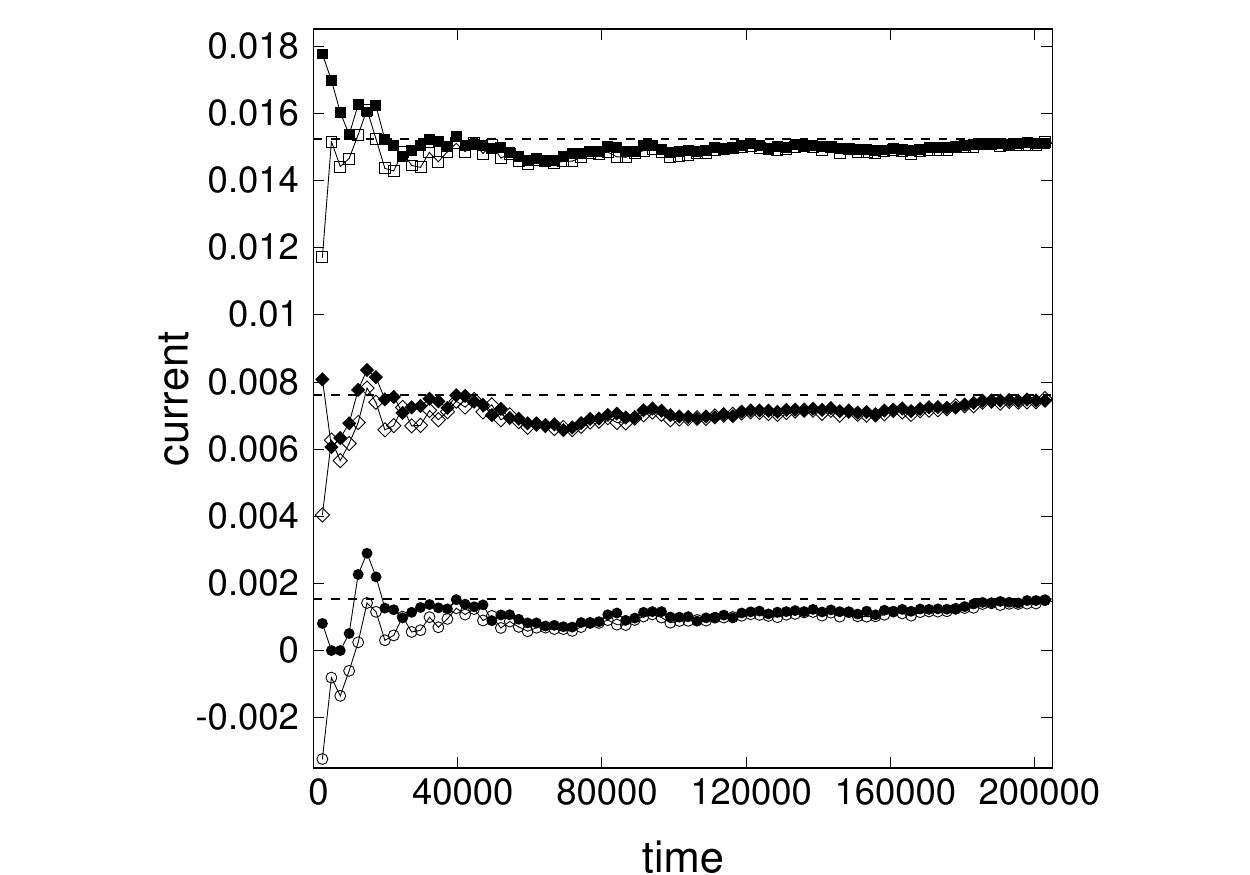}}%
}
\caption{
Results for the almost everywhere symmetric ZRP--BB model
with $R=50$, $N=206$, $\lambda=0.25$, $\epsilon=0.05$ (circles),
$0.2$ (diamonds),
$0.4$ (squares).
Left panel:
stationary density profiles, obtained as discussed in the caption of Fig. \ref{fig05}.
Open symbols denote the Monte Carlo measures and the dashed lines the 
exact solution in \eqref{cir100} and \eqref{cir110}. Note that the average occupation numbers of the \textit{finite reservoirs} in $x=0$ and in $x=2R+2$ have been multiplied by the factor $\lambda$ to fit within the figure.
Right panel: particle currents measured on the bonds 
$(2R+1)$--$(2R+2)$ (open symbols)
and 
$(2R+2)$--$(0)$ (solid symbols).
The dashed lines represent the theoretical prediction in \eqref{cir140}. Time, on the horizontal axis, is counted from the ``thermalization'' time $2\times10^6$, as in Figure \ref{fig04}.
}
\label{fig06}%
\end{figure*}

In this periodic case it is not possible to push forward the discussion 
without embracing a specific form for the intensity function. 
Then, from now onwards in this section,
we shall restrict our description to the independent particle case $u(k)=k$ and add the superscript ``ip" to the notation.
We first compute the partition function
\begin{equation}
\label{cir080}
Z^\textrm{ip}_{R,N}
=
\sum_{n\in\Omega_{R,N}}
\prod_{x=0}^{2R+2}
\frac{s_x^{n_x}}{n_x!}
=
\frac{1}{N!}\Big[\sum_{x=0}^{2R+2}s_x\Big]^N
\end{equation}
where we used the convention $0!=1$ and 
applied the multinomial theorem \cite[equation~(3.35)]{N1995}.
From \eqref{cir070} (and from the notational remark below it) we then get 
\begin{equation}
\label{cir090}
J^\textrm{ip}_{R,N}
=
N
\Big[\sum_{x=0}^{2R+2}s_x\Big]^{-1}
(p_xs_x-q_{x+1}s_{x+1})
\;\;.
\end{equation}
Moreover, since $u(n_x)=n_x$, the equation \eqref{cir060} can be 
also used to compute the density profile:
\begin{equation}
\label{cir100}
\rho^\textrm{ip}_x
=\nu_{R,N}^\textrm{ip}[n_x]
=\nu_{R,N}^\textrm{ip}[u(n_x)]
=
N
\Big[\sum_{x=0}^{2R+2}s_x\Big]^{-1}
s_x
\;\;,
\end{equation}
where in the last step we used \eqref{cir060} and \eqref{cir080}.
Note that $\rho^\textrm{ip}_0$ and 
$\rho^\textrm{ip}_{2R+2}$  
correspond to the average total number of particles allocated, respectively, in the left and 
in the right reservoirs. Retaining the interpretation discussed at the end of Sec. \ref{s:aperto}, from the expression of the injection rates given in \eqref{per010} and \eqref{per015} we find that the average particle densities in the two reservoirs take the values, respectively,  $\lambda \rho^\textrm{ip}_0$ and $\lambda \rho^\textrm{ip}_{2R+2}$. Thus, in the ZRP--CC model, the two slow sites act as finite particle reservoirs, each constituted by $\lambda^{-1}$ sites. In Figure \eqref{fig06} (right panel) 
shown are, hence, the density profile in the bulk, i.e. for $x=1,...,2R+1$, and at the slow sites in $x=0$ and in $x=2R+2$.

\subsection{The almost everywhere symmetric ZRP--CC}
\label{s:per-asy} 
\par\noindent
Let us fix 
$q=p=\gamma=\beta=1/2$,
$\bar{q}=1/2-\epsilon$, and 
$\bar{p}=1/2+\epsilon$ for some $\epsilon\in[0,1/2)$.
In this case the solution of the equations \eqref{cir030} is linear 
in $x$ and can be written as 
\begin{widetext}
\begin{equation}
\label{cir110}
s_x=
\left\{
\begin{array}{ll}
\sigma x+2\lambda s_0
& \textrm{ for }x=1,\dots,R
\\
2\lambda(3+2R)s_0/(3+2R-2\epsilon)
& \textrm{ for }x=R+1
\\
\sigma x
+2\lambda[1+4\epsilon+8\epsilon^2/(3+2R-2\epsilon)]s_0
& \textrm{ for }x=R+2,\dots,2R+1
\\
\left[1+4\epsilon /(3+2R-2\epsilon)\right]s_0
& \textrm{ for }x=2R+2
\\
\end{array}
\right.
\end{equation}
\end{widetext}
with slope 
\begin{equation}
\label{cir120}
\sigma=-\frac{8\lambda\epsilon s_0}{3+2R-2\epsilon}
\end{equation}
and $s_0$ arbitrary.
Moreover, we have that 
\begin{equation}
\label{cir130}
\sum_{x=0}^{2R+2}s_x
=
\frac{2(3+2R)[1+\lambda(2R+1)]}{3+2R-2\epsilon}s_0
\;\;.
\end{equation}
Hence, \eqref{cir090} and \eqref{cir100} yield
\begin{equation}
\label{cir140}
\begin{array}{rcl}
J^\textrm{ip}_{R,N}
&\!=&\!
{\displaystyle
-
\frac{1}{2}\sigma
N
\Big[\sum_{x=0}^{2R+2}s_x\Big]^{-1}
\vphantom{\bigg\{_\Big\}}
}
\\
&\!=&\!
{\displaystyle
\frac{2\lambda\epsilon N}{(3+2R)[1+\lambda(2R+1)]}
}
\\
\end{array}
\end{equation}
and 
\begin{displaymath}
\label{cir150}
\rho^\textrm{ip}_0
=
\frac{N(3+2R-2\epsilon)}{2(3+2R)[1+\lambda(2R+1)]}
\end{displaymath}
and
\begin{displaymath}
\rho^\textrm{ip}_{2R+2}
=
\frac{N(3+2R+2\epsilon)}{2(3+2R)[1+\lambda(2R+1)]}
\;\;.
\end{displaymath}
In conclusion, 
\begin{displaymath}
\rho^\textrm{ip}_{2R+2}
-
\rho^\textrm{ip}_0
=
\frac{2\epsilon N}{(3+2R)[1+\lambda(2R+1)]}
>0
\end{displaymath}
and
$J^\textrm{ip}_{R,N}>0$,
which proves that the channel is crossed by an uphill current 
flowing from the reservoir with lower particle density (in $x=0$)
to the one with higher particle density (in $x=2R+2$).

We have numerically simulated the almost everywhere symmetric ZRP--CC model following 
a scheme similar to that outlined in Section~\ref{s:ap-asy}.
We find, also in this case, an optimal match between the 
exact density profiles obtained from \eqref{cir100} and \eqref{cir110} and the 
numerical data, see Figure~\ref{fig06}.

\section{The hydrodynamic limit}
\label{s:idro} 
\par\noindent
We discuss on heuristic grounds 
the hydrodynamic limit 
\cite{DMP1991,KL1999}
of the almost everywhere symmetric ZRP--OB model introduced in Section~\ref{s:ap-asy},
with the intensity function corresponding to the 
independent particle case, namely $u(k)=k$.

For any $i\in\Lambda$ set
$x_i=i/(2R+1)$ so that 
$x_i\in[1/(2R+1),1]$. 
Denote by $n_i(t)$ the time--dependent \emph{density profile} at time $t$, i.e. $n_i(t)$ is the average number of particles occupying the site $i$ at time $t$. 
The change of the number 
of particles at a site in the bulk, i.e. 
$i\neq1,R,R+1,R+2,2R+1$, in a small interval $\Delta t$, can be estimated as
\begin{displaymath}
n_i(t+\Delta t)-n_i(t)
\!
=
\!
-n_i(t)\Delta t
+\frac{1}{2}n_{i-1}(t)\Delta t
+\frac{1}{2}n_{i+1}(t)\Delta t
\end{displaymath}
This equality can be rewritten as
\begin{displaymath}
\frac{n_i(t+\Delta t)-n_i(t)}{\Delta t/(2R+1)^2}
\!
=
\!
\frac{[n_{i+1}(t)-n_i(t)]-[n_i(t)-n_{i-1}(t)]}
     {2/(2R+1)^2}
\end{displaymath}
Thus, if time is rescaled as $t/(2R+1)^2\rightarrow t$ (diffusive scaling), in the limit $R\to\infty$ 
the particle density profile $n_i(t)$ will tend to a function 
$u(x,t)$ solving the diffusion equation 
\begin{equation}
\label{idro000}
\frac{\partial u}{\partial t}
=
\frac{1}{2}
\frac{\partial^2 u}{\partial x^2}
\;\;\textrm{ in }
(0,1/2)\cup(1/2,1)
\;\;.
\end{equation}

In order to 
guess the boundary conditions at $x=0,1/2,1$ we shall write the balance equation
of the currents at the sites $x_1$, 
$x_{R}$, $x_{R+1}$, $x_{R+2}$, and $x_{2R+1}$.
More precisely, we consider a small interval of time $\Delta t$ and 
we first write
\begin{displaymath}
\alpha\Delta t-n_1(t)\Delta t+\frac{1}{2}n_2(t)\Delta t=0
\end{displaymath}
and
\begin{displaymath}
\delta\Delta t-n_{2R+1}(t)\Delta t+\frac{1}{2}n_{2R}(t)\Delta t=0
\end{displaymath}
which, in the limit $R\to\infty$, provide the boundary conditions
\begin{equation}
\label{idro010}
u(0,t)=2\alpha
\;\;\textrm{ and }\;\;
u(1,t)=2\delta
\;\;.
\end{equation}
Note that Eqs. \eqref{idro010} are obtained by assuming that the 
injection rates $\alpha$ and $\delta$ are independent of $R$;
different boundary conditions may hold under different scalings of 
$\alpha$ and $\delta$ with $R$.
Moreover, we have that
\begin{displaymath}
\begin{array}{l}
{\displaystyle
 \frac{1}{2}n_{R-1}(t)\Delta t
 -n_R(t)\Delta t
 +\Big(\frac{1}{2}-\epsilon\Big)n_{R+1}(t)\Delta t
}
=
0
\vphantom{\bigg\{_\}}
\\
{\displaystyle
 \frac{1}{2}n_{R}(t)\Delta t
 -n_{R+1}(t)\Delta t
 +\frac{1}{2}n_{R+2}(t)\Delta t
}
=
0
\vphantom{\bigg\{_\}}
\\
{\displaystyle
 \Big(\frac{1}{2}+\epsilon\Big)n_{R+1}(t)\Delta t
 -n_{R+2}(t)\Delta t
 +\frac{1}{2}n_{R+3}(t)\Delta t
}
=
0
\end{array}
\end{displaymath}
The equation in the middle can be rewritten as follows
\begin{displaymath}
 \frac{1}{2}(n_{R}(t)-n_{R+1}(t))
 =
 \frac{1}{2}(n_{R+1}(t)-n_{R+2}(t))
\end{displaymath}
which, divided by $1/(2R+1)$, in the limit $R\to\infty$ provides 
the condition
\begin{equation}
\label{idro020}
\lim_{x\to 1/2^-}\frac{\partial}{\partial x}u(x,t)
=
\lim_{x\to 1/2^+}\frac{\partial}{\partial x}u(x,t)
\;\;.
\end{equation}
Combining the first and the third equation, on the other hand, 
we get 
\begin{displaymath}
\begin{array}{l}
{\displaystyle
 \Big[\frac{1}{2}n_{R-1}(t)-n_R(t)\Big]\Big(\frac{1}{2}+\epsilon\Big)
 \vphantom{\Bigg\{_\Big\}}
}
\\
\phantom{mmmm}
+
{\displaystyle
 \Big[n_{R+2}(t)-\frac{1}{2}n_{R+3}(t)\Big]\Big(\frac{1}{2}-\epsilon\Big)
 =0\; .
}
\end{array}
\end{displaymath}
Since in the limit $R\to\infty$ we have 
that
$[n_{R-1}(t)-n_R(t)]/2$
and
$[n_{R+2}(t)-n_{R+3}(t)]/2$
tend to zero, the above equation can be interpreted as 
\begin{equation}
\label{idro030}
\Big(\frac{1}{2}+\epsilon\Big)
\lim_{x\to 1/2^-}u(x,t)
=
\Big(\frac{1}{2}-\epsilon\Big)
\lim_{x\to 1/2^+}u(x,t)
\;\;.
\end{equation}

In conclusion, we find that the evolution of the model in the 
hydrodynamic limit is described by the differential equation
\eqref{idro000} supplemented with the boundary conditions 
\eqref{idro010}, \eqref{idro020}, and \eqref{idro030}.
In particular, the stationary profile is the solution of the problem
\begin{equation}
\label{idro040}
\left\{
\begin{array}{l}
u''(x)=0
\\
u(0)=2\alpha\;\textrm{ and }\;u(1)=2\delta
\\
u'_-(1/2)=u'_+(1/2)
\\
(1/2+\epsilon)u_-(1/2)=(1/2-\epsilon)u_+(1/2)
\end{array}
\right.
\end{equation}
where the subscripts $-$ and $+$ denote, respectively, the left and the right limits.

\begin{figure*}[t]
\centerline{%
\hspace{-.5 cm}
{\includegraphics[width=0.35\textwidth]{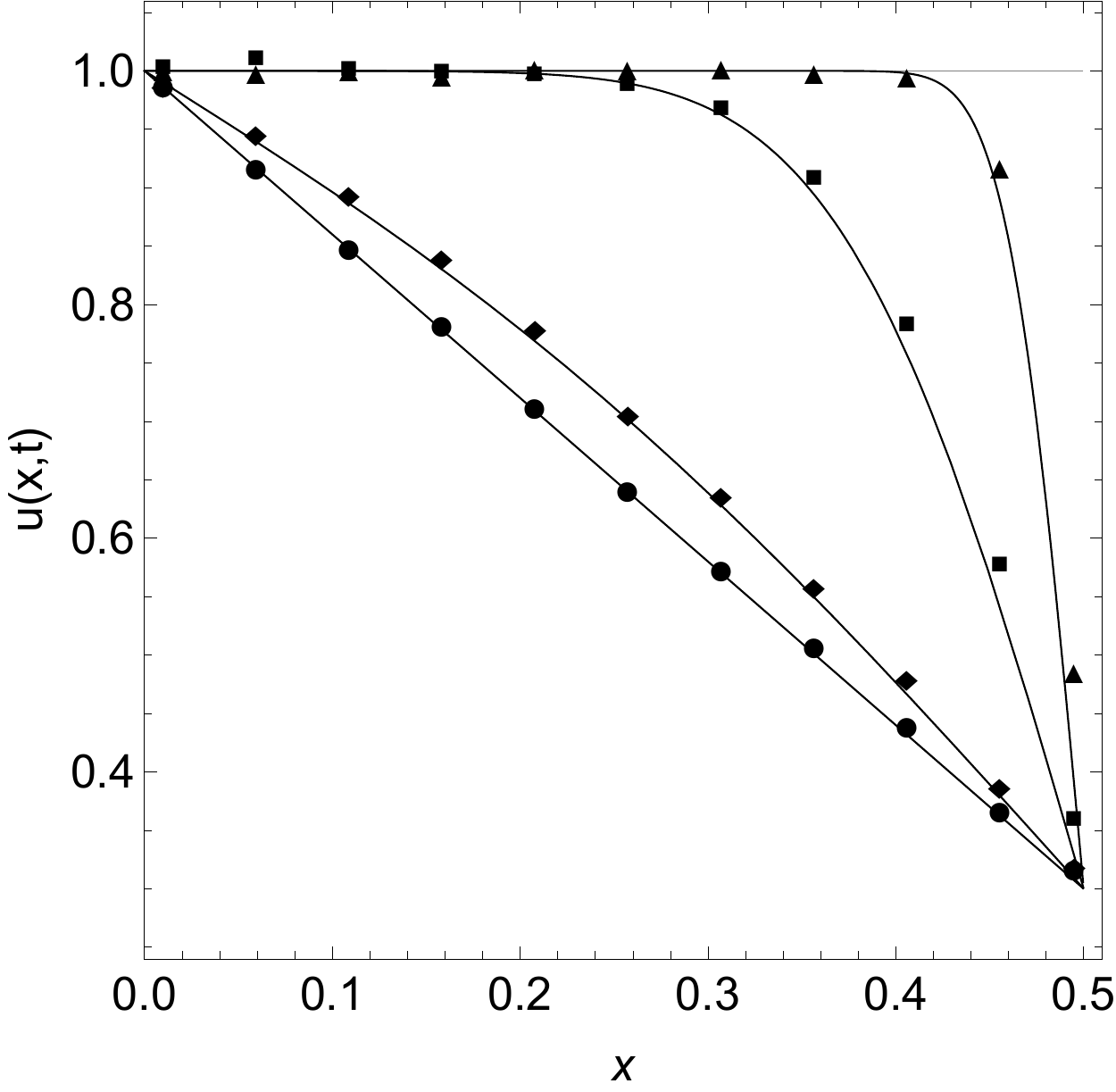}}%
\hspace{1.5 cm}
{\includegraphics[width=0.35\textwidth]{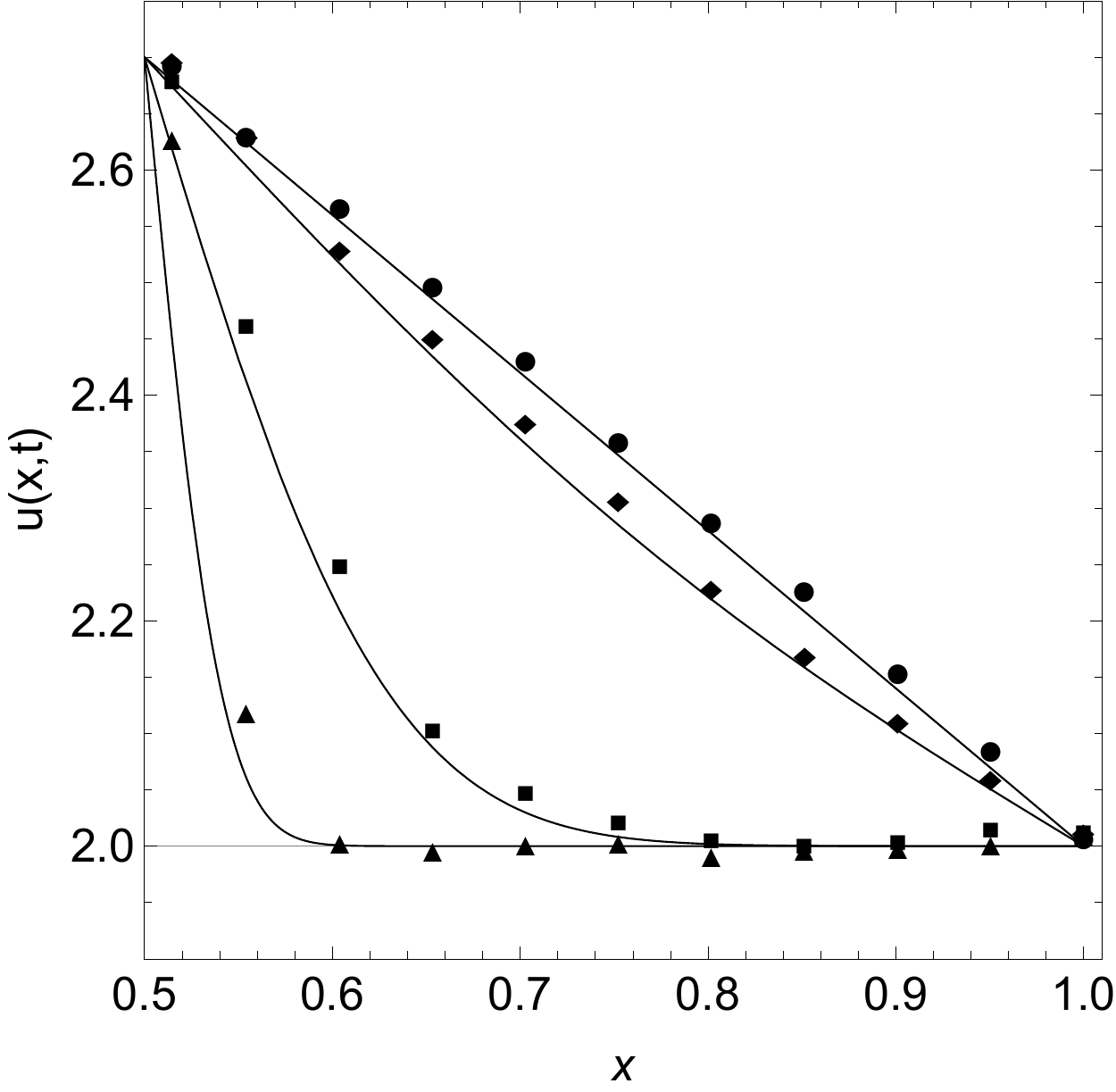}}%
}
\caption{Comparison between the exact solution of the hydrodynamic 
problem \eqref{idro060}--\eqref{idro090} and the Monte Carlo 
measure of the time--dependent density profiles, averaged over a set of different realizations of the stochastic process. 
The parameters of the simulation are $\alpha=0.5$, $\delta=1$, $\epsilon=0.4$ 
and $R=50$.
The profiles are plotted at times $t=0.001$ (triangles),
$t=0.01$ (squares), $t=0.1$ (diamonds), and $t=0.5$ (circles).
The gray and the black solid lines denote, respectively, the initial condition and 
the exact solution at the corresponding times.
The profiles in the regions $[0,1/2]$ and $[1/2,1]$ are displayed 
in two separate panels to optimize the resolution of the plots. 
}
\label{fig07}%
\end{figure*}

The stationary problem \eqref{idro040} can be easily solved:
one can write $u(x)=Ax+B$ for $x\in(0,1/2)$ and 
$u(x)=Cx+D$ for $x\in(1/2,1)$. The boundary conditions then yield
\begin{equation}
\label{idro050}
u(x)=
2[\delta-\alpha-2\epsilon(\alpha+\delta)]x+2\alpha
\end{equation}
for $0\le x\le1/2$
and
\begin{equation}
\label{idro051}
u(x)=
2[\delta-\alpha-2\epsilon(\alpha+\delta)]x+2\alpha+4\epsilon(\alpha+\delta)
\end{equation}
for $1/2\le x\le1$.
The solution of the macroscopic stationary equation matches perfectly 
with the stationary density profile of the microscopic lattice model. 
Indeed, by performing the change of variable $x/(2R+1)\to x$ in 
\eqref{uph100}, one finds, for $R$ large, the equations \eqref{idro050} and \eqref{idro051}. 

It is also possible to solve the time dependent 
problem \eqref{idro000}--\eqref{idro030}
and write the solution in terms of a Fourier series. 
We first introduce the functions 
\begin{displaymath}
Y_1(x,t)=u(x,t)
\;\;\textrm{ and }\;\;
Y_2(x,t)=u(1-x,t)
\end{displaymath}
for $x\in[1,1/2]$
and note that the conditions \eqref{idro010}--\eqref{idro030}
imply
\begin{displaymath}
Y_1(0,t)=2\alpha,
Y_2(0,t)=2\delta,
\frac{\partial Y_1}{\partial x}\Big(\frac{1}{2},t\Big)
+
\frac{\partial Y_2}{\partial x}\Big(\frac{1}{2},t\Big)
=0\;,
\end{displaymath}
and 
\begin{displaymath}
\Big(\frac{1}{2}+\epsilon\Big)Y_1(1/2,t)
-
\Big(\frac{1}{2}-\epsilon\Big)Y_2(1/2,t)
=0
\;\;.
\end{displaymath}

\begin{figure*}[t]
\centerline{%
\hspace{-.5 cm}
{\includegraphics[width=0.35\textwidth]{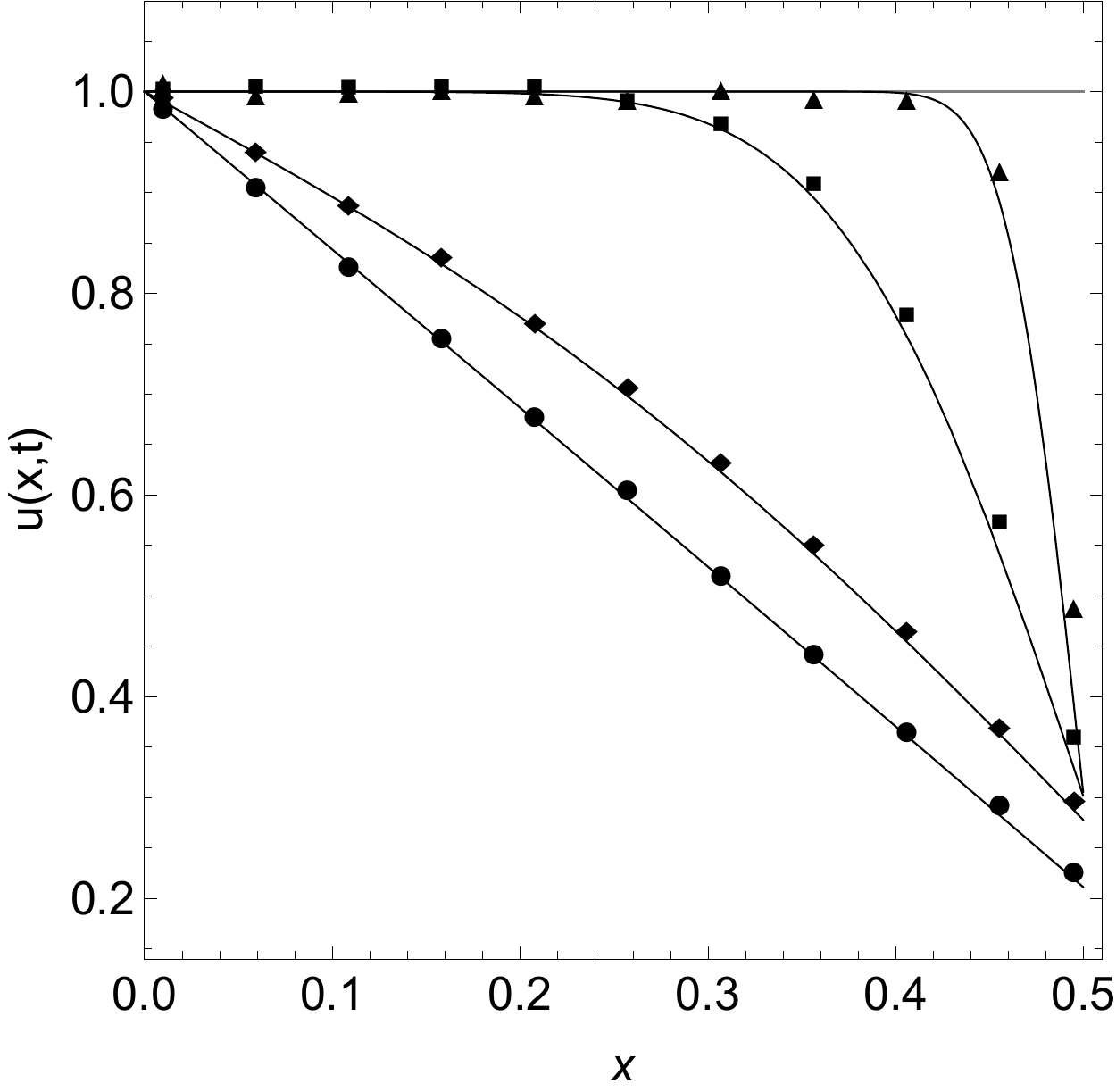}}%
\hspace{1.5 cm}
{\includegraphics[width=0.35\textwidth]{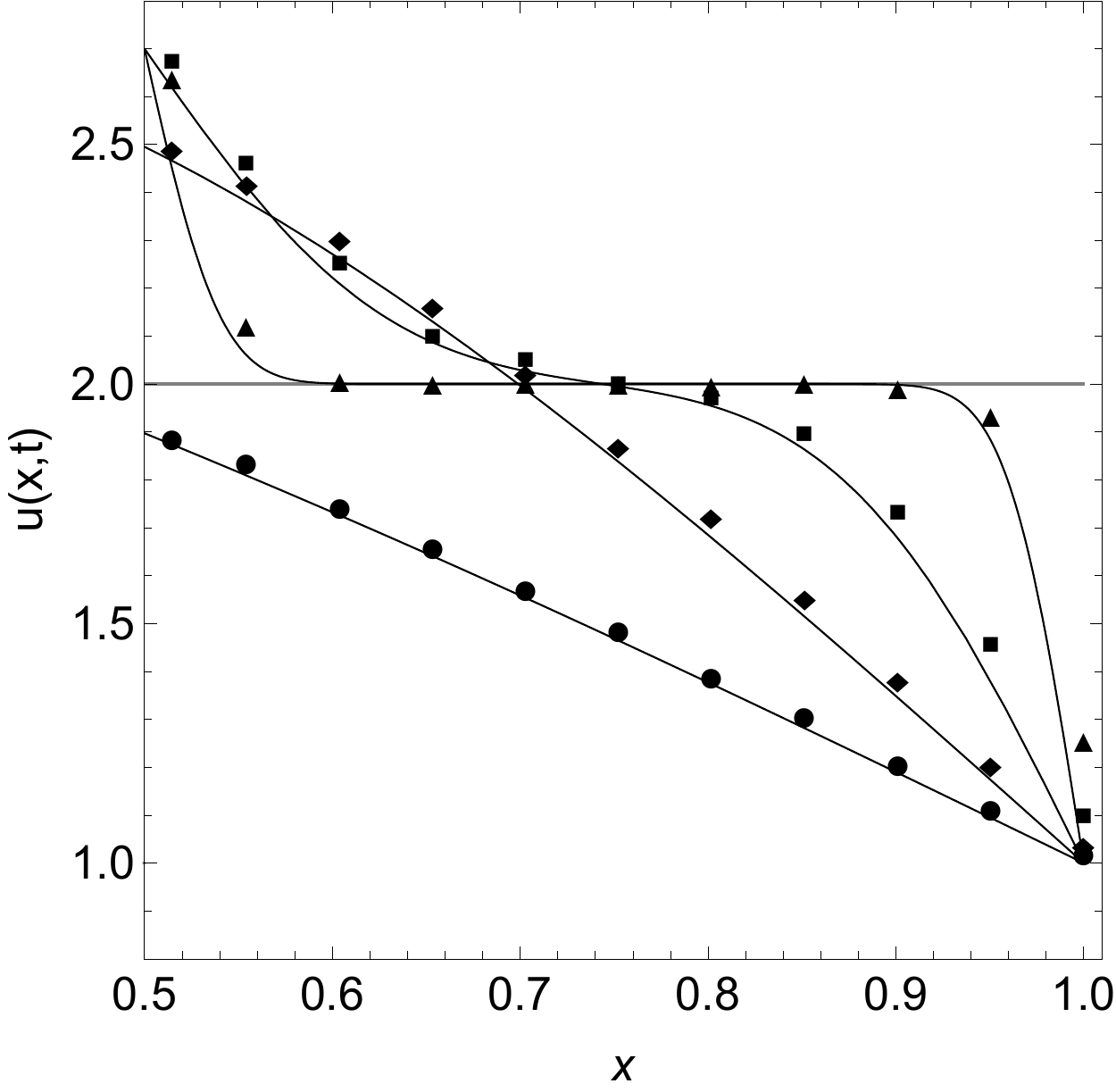}}%
}
\caption{Comparison between the exact solution of the hydrodynamic 
problem \eqref{idro060}--\eqref{idro090} and the Monte Carlo 
measure of the time--dependent density profiles, with $\alpha=\delta=0.5$, $\epsilon=0.4$ 
and $R=50$. Symbols are the same as those shown in Figure \ref{fig07}.}
\label{fig08}%
\end{figure*}

Moreover, by \eqref{idro000} we have that both $Y_1$ and $Y_2$ 
solve the heat equation with diffusion coefficient $1/2$.
As a second step, we introduce the functions 
\begin{displaymath}
W(x,t)=Y_1(x,t)+Y_2(x,t)
\end{displaymath}
and
\begin{displaymath}
U(x,t)
=
\Big(\frac{1}{2}+\epsilon\Big)Y_1(x,t)
-
\Big(\frac{1}{2}-\epsilon\Big)Y_2(x,t)
\end{displaymath}
and note that from the boundary conditions on $Y_1$ and $Y_2$ we get 
\begin{displaymath}
W(0,t)=2(\alpha+\delta),\;
W(1/2,t)=2(\alpha+\delta)\;,
\end{displaymath}
and
\begin{displaymath}
U(0,t)=\alpha-\delta+2\epsilon(\alpha+\delta),\;
U(1/2,t)=0\;.
\end{displaymath}

Thus, we obtained two PDE problems, one for $W$ and another for $U$, which 
are decoupled and can hence be solved by the standard method of separation 
of variables. 
Denoting by $u_0(x)$ the initial condition for the original 
equation \eqref{idro000}, we can define 
\begin{displaymath} 
Y_{1,0}(x)=u_0(x)
\;\;\textrm{ and }\;\;
Y_{2,0}(x)=u_0(1-x)
\end{displaymath} 
for $x\in[0,1/2]$.
Moreover, we set 
\begin{displaymath} 
W_0(x)=Y_{1,0}(x)+Y_{2,0}(x)
\end{displaymath} 
and
\begin{displaymath} 
U_0(x)
=
\Big(\frac{1}{2}+\epsilon\Big)Y_{1,0}(x)
-
\Big(\frac{1}{2}-\epsilon\Big)Y_{2,0}(x)
\;\;.
\end{displaymath} 

Then, by a standard computation, we find 
\begin{equation}
\label{idro060}
W(x,t)=2(\alpha+\delta)
+\sum_{n=0}^\infty
A_n
e^{-\alpha_n^2t/2}\sin(\alpha_nx)
\end{equation}
with 
$\alpha_n=(1+2n)\pi$
and
\begin{displaymath}
A_n=4\int_0^{1/2}[W_0(x)-2(\alpha+\delta)]\sin(\alpha_nx)\,\textrm{d}x
\end{displaymath}
for $n=0,1,\dots$.
For the function $U$ we find
\begin{equation}
\label{idro070}
\begin{array}{l}
U(x,t)=(1-2x)[\alpha-\delta+2\epsilon(\alpha+\delta)]
\\
{\displaystyle
\phantom{mmmmi}
+\sum_{n=0}^\infty
B_n
e^{-\beta_n^2t/2}\sin(\beta_nx)
}
\end{array}
\end{equation}
with 
$\beta_n=2n\pi$
and
\begin{displaymath}
B_n=4\int_0^{1/2}\!\!\!\!\!
[U_0(x)-(1-2x)[\alpha-\delta+2\epsilon(\alpha+\delta)]
 \sin(\beta_nx)]\,\textrm{d}x
\end{displaymath}
for $n=0,1,\dots$.
Solving the equations that define $W$ and $U$ with respect to 
$Y_1$ and $Y_2$ we find 
\begin{displaymath}
Y_1(x,t)=\Big(\frac{1}{2}-\epsilon\Big)W(x,t)+U(x,t)
\end{displaymath}
and
\begin{displaymath}
Y_2(x,t)=\Big(\frac{1}{2}+\epsilon\Big)W(x,t)-U(x,t)
\;\;.
\end{displaymath}
Finally, we get the solution of the original 
problem, 
\begin{equation}
\label{idro090}
u(x,t)
=
\left\{
\begin{array}{ll}
Y_1(x,t) & \textrm{ for } x\in[0,1/2]\\
Y_2(1-x,t) & \textrm{ for } x\in[1/2,1]\;\;.\\
\end{array}
\right.
\end{equation}

\begin{figure*}[t]
\centerline{%
\hspace{-.5 cm}
{\includegraphics[width=0.35\textwidth]{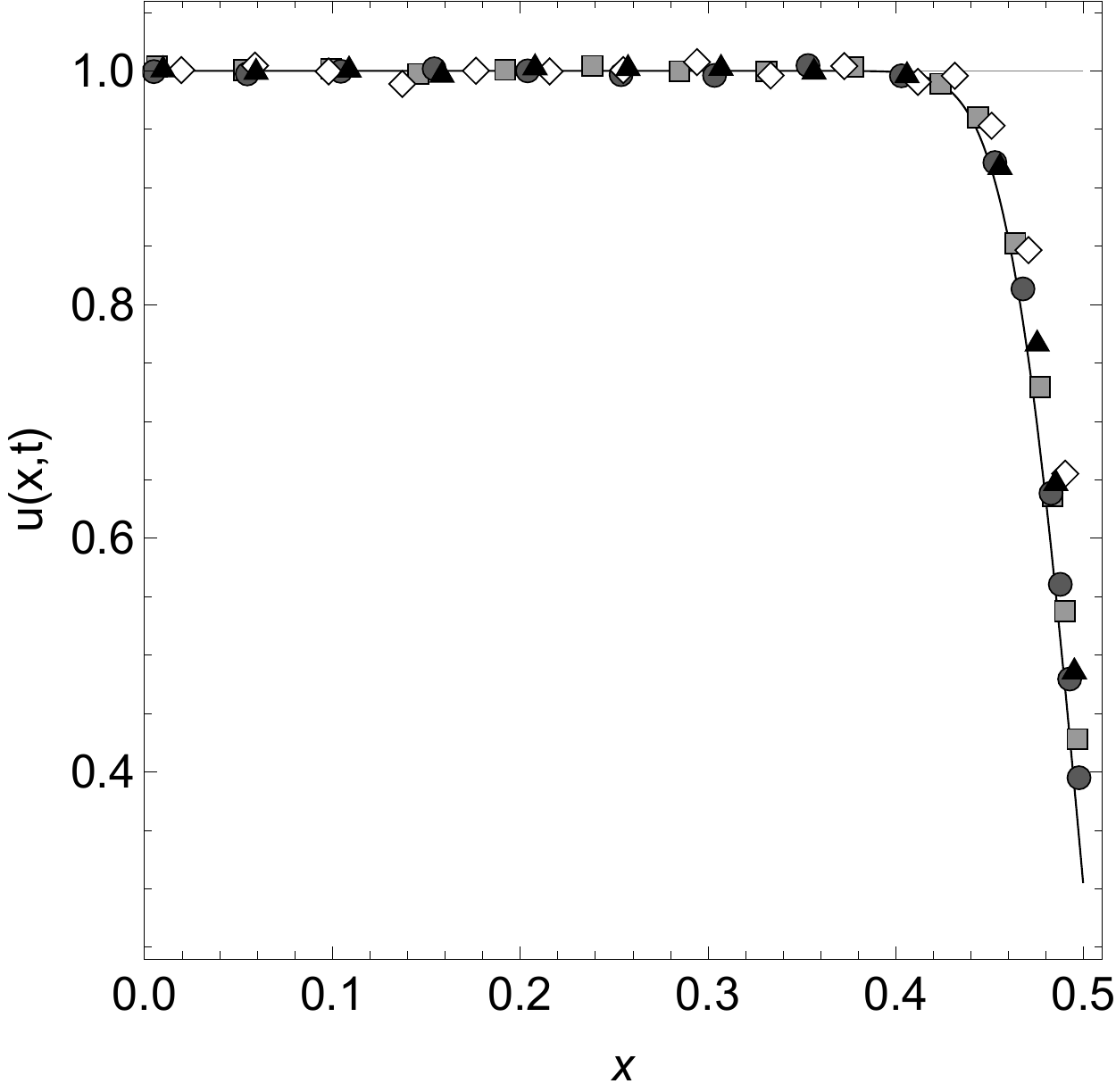}}%
\hspace{1.5 cm}
{\includegraphics[width=0.35\textwidth]{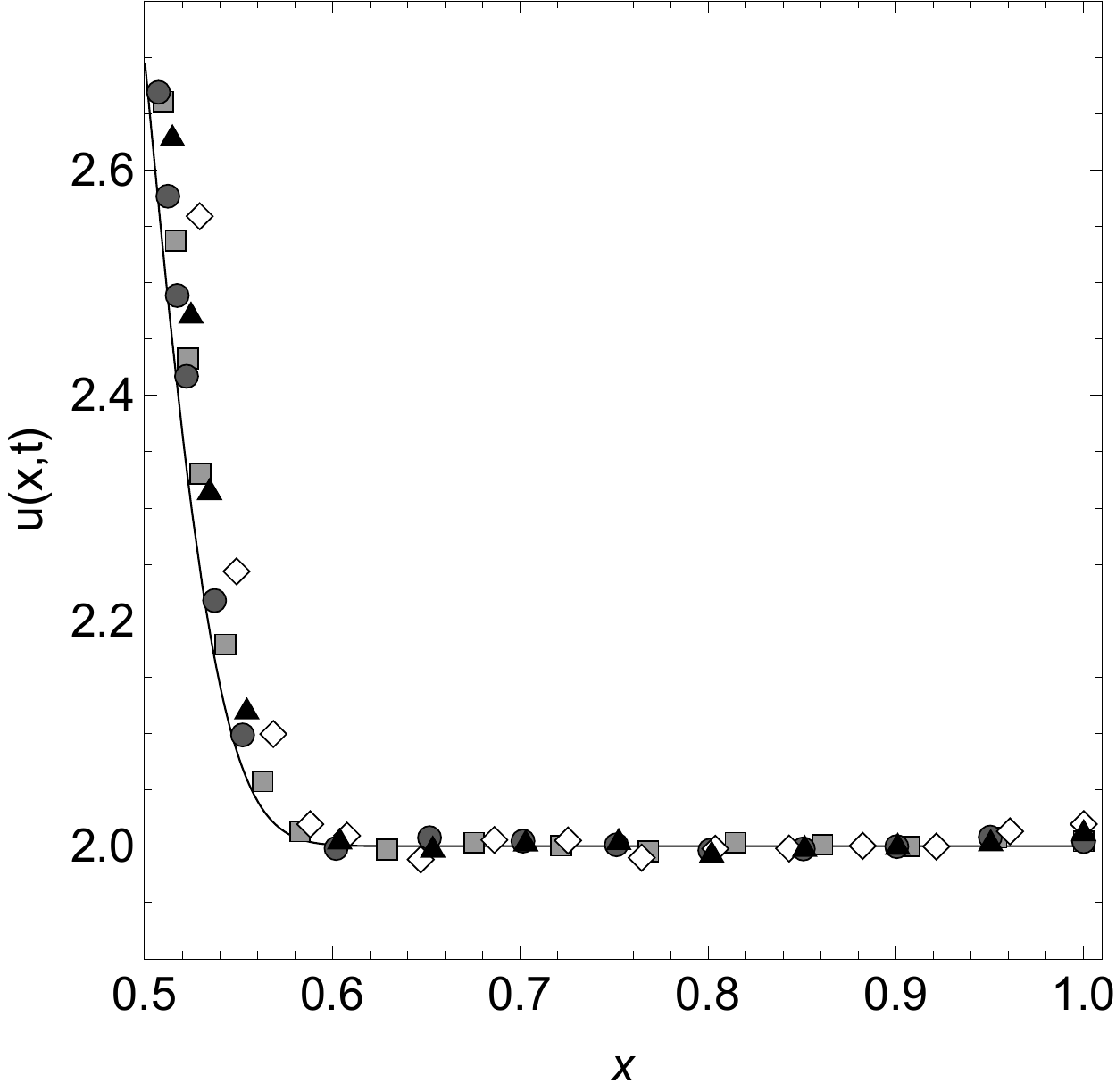}}%
}
\caption{Comparison between the exact solution of the hydrodynamic 
problem \eqref{idro060}--\eqref{idro090} (black solid lines) and the Monte Carlo 
measure, with $\alpha=0.5$, $\delta=1.0$, $\epsilon=0.4$, at time 
$t=0.001$, for different volume sizes: $R=25$ (empty diamonds), 
$R=50$ (black triangles), $R=75$ (gray squares) and $R=100$ (dark gray circles). The gray solid lines denote the initial condition.}
\label{fig09}%
\end{figure*}

We now test numerically the solution \eqref{idro060}--\eqref{idro090}.
We consider the hydrodynamic problem in the case 
$\alpha=0.5$ and $\delta=1$ in Figure~\ref{fig07} 
and 
$\alpha=\delta=0.5$ in Figure~\ref{fig08}. 
In both figures $\epsilon=0.4$ and the initial datum is 
$u_0(x)=1$ for $x\in[0,1/2]$ and 
$u_0(x)=2$ for $x\in[1/2,1]$.
The density profile is plotted at different macroscopic times 
and is compared with the numerical estimate. 

The numerical solution is constructed as follows: a set of $5\times 10^5$  
independent realizations of the stochastic process is constructed by 
running different Monte Carlo simulations started from the same initial 
datum (the one also used for the analytical solution) 
and by varying the seed of the random number generator routine. 
Then, the profile corresponding to a certain fixed macroscopic time is obtained by averaging over all the different realizations of the process. 
Finally, the numerical profile is plotted after rescaling 
the space microscopic variable as 
$x/(2R+1)\to x$ and the very good match illustrated in 
Figures~\ref{fig07} and \ref{fig08} is found. 

It should be observed that, in both Figures~\ref{fig07} and \ref{fig08}, the Monte Carlo results display some (little) discrepancies with respect to the theoretical behavior indicated by the solid lines. These are fluctuations stemming from finite size effects. 

Indeed, fixed the initial datum,
averaging over a (large enough) set of different realizations of 
the process corresponds to considering the expectation 
$
\mathbb{E}_{\mu_R^t}[n_{x}(t)]
$ 
with respect to a probability measure $\mu_R^t$ associated with the 
stochastic process at time $t$.
We recall, then, that the  hydrodynamic behavior holds in the 
limit $R\rightarrow \infty$.
More precisely, one introduces the \textit{empirical density} \cite{BDSGJL02}
\begin{equation}
\label{empm}
\pi_R^t(n)=\frac{1}{2R+1}\sum_{x\in \Lambda}n_x(t)\delta_x\;,
\end{equation}
where $\delta_x$ is the delta measure. From Eq. \eqref{empm} one finds that, for
any continuous function $f:\Omega_R\to \mathbb{R}$, it holds
$$
\int_{\Omega_R} f\,d\pi_R^t(n)=\frac{1}{2R+1}\sum_{x\in \Lambda}n_x(t)f(x)\;.
$$ 
One says, then, that a sequence of probability measures $\mu_R^t$ on $\Omega_R$ is associated with a density profile $u(x,t)$ if for any continuous function $f$ and for any $\epsilon >0$ it holds
\begin{equation}
\label{convprob}
\lim_{R\to+\infty}\mathbb{E}_{\mu_R^t}\left[\mathbf{1}_{\left|\int_{\Omega_R} f\,d
\pi_R^t(n)-\int_{\Omega_R} f(x)u(x,t)dx\right|\geq \epsilon}\right]=0\,,\nonumber\\
\end{equation}
where $\mathbf{1}$ denotes the characteristic function.

In Figure~\ref{fig09} we show that the match between the solution of the 
hydrodynamic limit equations and the numerical simulation becomes better 
and better when the size of the lattice used in the simulations increases. 
The same situation as the one portrayed in 
Figure~\ref{fig07} at the macroscopic time 
$0.001$ is considered and simulations are run for 
$R=25, 50, 75,100$. Note that the case with $R=50$ (black triangles) is also the case 
shown in 
Figure~\ref{fig07}. 

\section{Conclusions}
\label{s:conclusioni} 
\par\noindent
A variety of systems, e.g. two--species models, particle or spin models undergoing a phase transition, queuing network models, are known to exhibit uphill currents.
In this paper we prove that the phenomenon of uphill diffusion can also be observed in the 
simplest and, somehow, paradigmatic transport model, namely the 
1D Zero Range Process. 

Indeed, such a model is proven to 
show uphill currents in presence of a bias on a single 
defect site. For an open ZRP in contact with two particle reservoirs 
at different densities, for sufficiently large volumes the density at the boundaries of the channel 
depends only on the injection rates and not on the local bias. If the bias is large enough the current changes sign, so that particles typically move 
uphill, from the reservoir with lower density to the one with higher density. 
This result is demonstrated both analytically and numerically, with a striking match 
between the exact and the Monte Carlo results. 

We have also investigated the hydrodynamic limit of the model: a heuristic 
argument yields the structure of the limit problem and provides the matching conditions mimicking the presence  of the defect site in the microscopic lattice model. 
We managed to write the time dependent solution as a Fourier series 
and compared it with the evolution of the original ZRP process.

\begin{acknowledgments}
We thank A.\ De Masi and E.\ Presutti for inspiring this work 
and for the many enlightening discussions. 
We also thank D.\ Andreucci and D.\ Gabrielli for useful discussions
on problems related to the derivation of hydrodynamic limits in presence 
of local discontinuities. 
\end{acknowledgments}


\end{document}